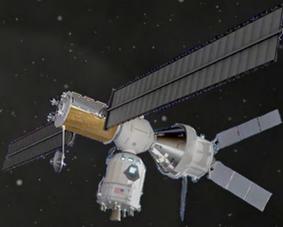

# FARSIDE

## Farside Array for Radio Science Investigations of the Dark ages and Exoplanets

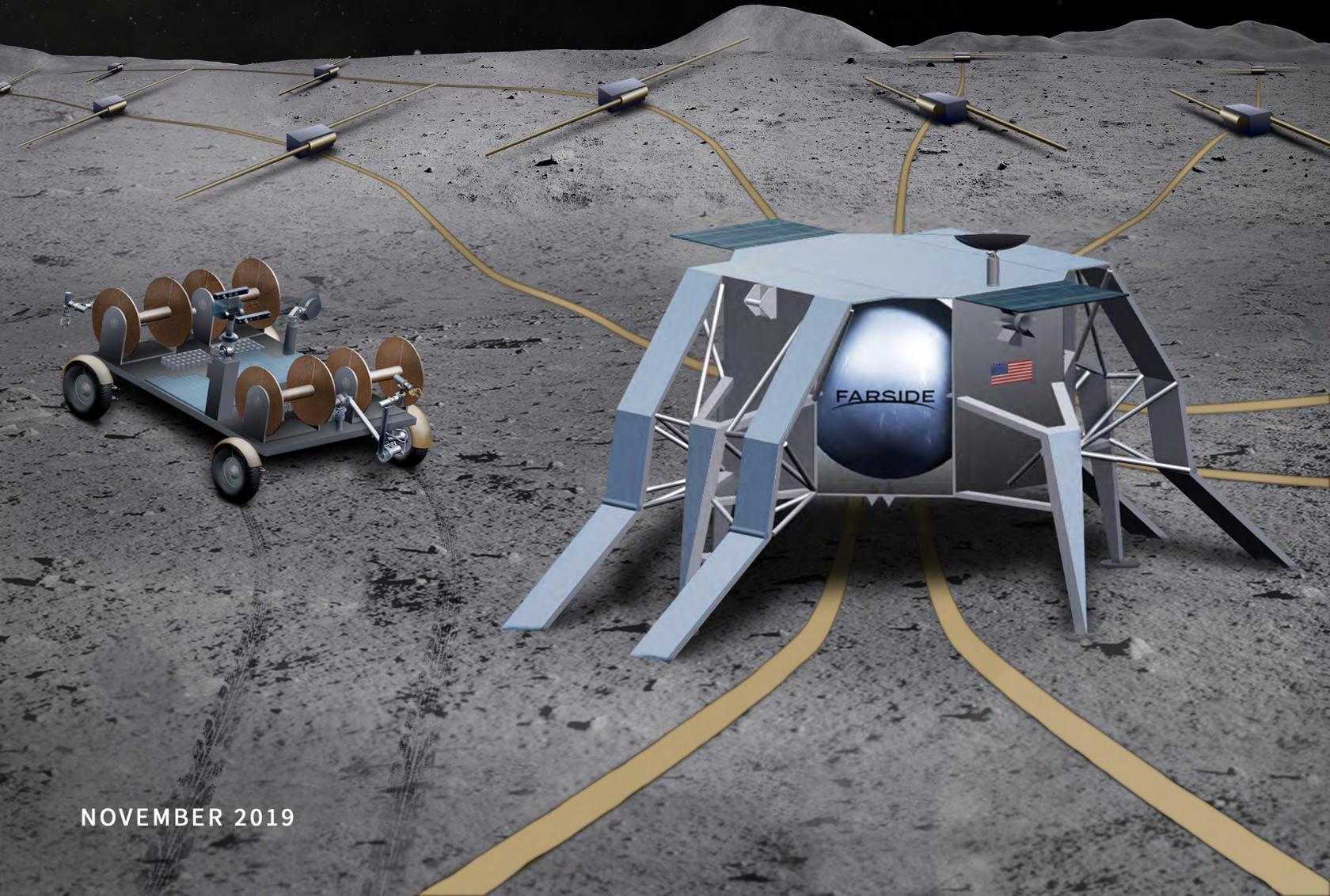

NOVEMBER 2019

## Study Participants List

| Principal Authors | |
|---|---|
| Jack O. Burns, University of Colorado Boulder | Gregg Hallinan, California Institute of Technology |
| **Co-Authors** | |
| Jim Lux, NASA Jet Propulsion Laboratory, California Institute of Technology | Andres Romero-Wolf, NASA Jet Propulsion Laboratory, California Institute of Technology |
| Lawrence Teitelbaum, NASA Jet Propulsion Laboratory, California Institute of Technology | Tzu-Ching Chang, NASA Jet Propulsion Laboratory, California Institute of Technology |
| Jonathon Kocz, California Institute of Technology | Judd Bowman, Arizona State University |
| Robert MacDowall, NASA Goddard Space Flight Center | Justin Kasper, University of Michigan |
| Richard Bradley, National Radio Astronomy Observatory | Marin Anderson, California Institute of Technology |
| David Rapetti, University of Colorado Boulder | Zhongwen Zhen, California Institute of Technology |
| Wenbo Wu, California Institute of Technology | Jonathan Pober, Brown University |
| Steven Furlanetto, UCLA | Jordan Mirocha, McGill University |
| Alex Austin, NASA Jet Propulsion Laboratory, California Institute of Technology | |

## Disclaimers/Acknowledgements

Part of this research was carried out at the Jet Propulsion Laboratory, California Institute of Technology, under a contract with the National Aeronautics and Space Administration.

The cost information contained in this document is of a budgetary and planning nature and is intended for informational purposes only. It does not constitute a commitment on the part of JPL and/or Caltech

## Table of Contents

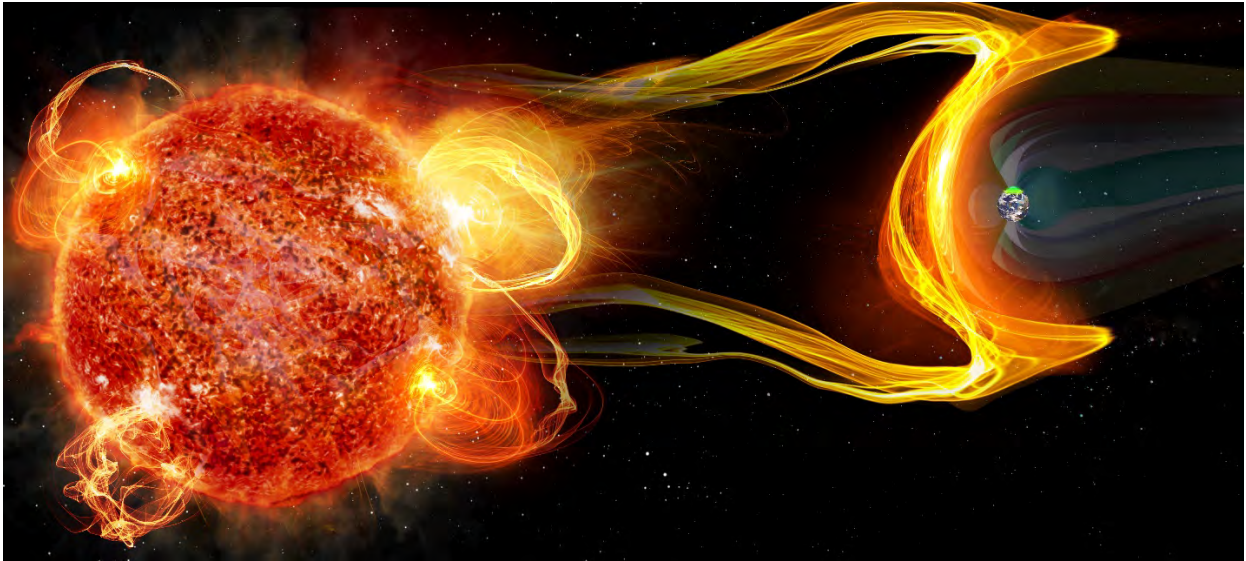

An artist's impression of a habitable planet experiencing a coronal mass ejection (CME) from its host star, an active M dwarf. Detecting coronal mass ejections, energetic particle events and the magnetospheres of candidate habitable planets is a key science goal for the FARSIDE array. Credit: Chuck Carter/Caltech/KISS.

# 1 EXECUTIVE SUMMARY

FARSIDE (*Farside Array for Radio Science Investigations of the Dark ages and Exoplanets*) is a Probe-class concept to place a low radio frequency interferometric array on the farside of the Moon. A NASA-funded design study, focused on the instrument, a deployment rover, the lander and base station, delivered an architecture broadly consistent with the requirements for a Probe mission. This notional architecture consists of 128 dipole antennas deployed across a 10 km area by a rover, and tethered to a base station for central processing, power and data transmission to the Lunar Gateway, or an alternative relay satellite. FARSIDE would provide the capability to image the entire sky each minute in 1400 channels spanning frequencies from 100 kHz to 40 MHz, extending down two orders of magnitude below bands accessible to ground-based radio astronomy. The lunar farside can simultaneously provide isolation from terrestrial radio frequency interference, auroral kilometric radiation, and plasma noise from the solar wind. It is thus the only location within the inner solar system from which sky noise limited observations can be carried out at sub-MHz frequencies. This would enable near-continuous monitoring of the nearest stellar systems in the search for the radio signatures of coronal mass ejections and energetic particle events, and would also detect the magnetospheres for the nearest candidate habitable exoplanets. Simultaneously, FARSIDE would be used to characterize similar activity in our own solar system, from the Sun to the outer planets, including the hypothetical Planet Nine. Through precision calibration via an orbiting beacon, and exquisite foreground characterization, FARSIDE would also measure the Dark Ages global 21-cm signal at redshifts z ~50–100. It will also be a pathfinder for a larger 21-cm power spectrum instrument by carefully measuring the foreground with high dynamic range. The unique observational window offered by FARSIDE would enable an abundance of additional science ranging from sounding of the lunar subsurface to characterization of the interstellar medium in the solar system neighborhood.

The FARSIDE instrument front-end uses electrically short simple dipole antennas to achieve sky background noise-limited observations. The regolith in the lunar highlands is thick, has low conductivity, and varies slowly with depth removing the need for a ground plane. Calibration will make use of astronomical sources, as well as an orbiting calibration beacon that will map each antenna gain pattern together with the effect of the lunar subsurface. The implementation uses thin wires on the ground with the rest of the system leveraging existing designs: high space heritage front-end amplifiers, fiber optics, and a correlator system used in ground-based radio astronomical observatories such as the Owens Valley Radio Observatory Long Wavelength Array (OVRO-LWA).

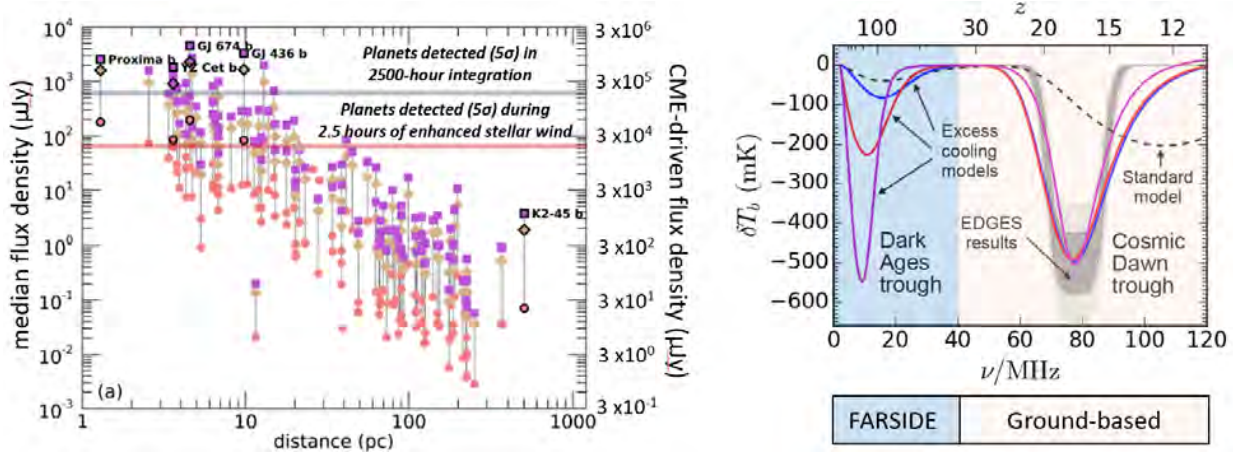

**Figure 1-1.** FARSIDE will pioneer advances in two key fields of astrophysics, identified as high science priorities in NASA's Science Plan and the Astro2010 Decadal Survey—exoplanets and high redshift cosmology. *Left:* The predicted average flux radio density at ~280 kHz of known exoplanets, orbiting M dwarfs [figure adapted from 5], assuming a magnetic field of 0.1 G (10% of Earth's magnetic field). Purple, yellow and red data points reflect different models for the quiescent stellar wind. The horizontal grey line represents the 5σ detection limit (y-axis on left-hand side) on the median flux density for the auroral radio emission detectable from FARSIDE in a 2500-hour integration. The horizontal red line represents the 5σ detection limit (y-axis on right-hand side) in 2.5 hours during enhanced stellar wind conditions, e.g., during a CME, assuming a similar degree of variability as observed for Earth's auroral kilometric radiation. *Right:* FARSIDE will be a sensitive probe of cosmology in the very early Universe. The black dashed curve shows the brightness temperature of neutral hydrogen in the standard cosmological model with adiabatic cooling. The shape at z > 30 is independent of astrophysical sources. The grey contours show the 1 and 2-sigma absorption bands inferred by recent EDGES results. The solid curves are parametric models that invoke extra cooling, possibly produced by dark matter, to match the amplitude in the EDGES signal but also dramatically affect the dark ages absorption trough at z > 50. FARSIDE will cleanly distinguish between the standard cosmology model with adiabatic expansion and models with added cooling, possibly due to new interactions with dark matter, at > 5-sigma level.

The FARSIDE mission includes three components: a commercial lander (e.g., part of NASA's Commercial Lunar Payload Services) carrying the FARSIDE base station and a rover which deploys the array of antenna nodes. The base station includes a pair of enhanced Multi Mission Radioisotope Thermoelectric Generators (eMMRTG) to provide power for the receivers and the signal processing functions of the FX correlator, as well as communicating with the Lunar Gateway (or a similar data relay capability from the lunar farside) to send the science data back to Earth. The rover will be solar powered, hibernating at night, and also includes a telecommunications subsystem to communicate with the Gateway during the array deployment operations phase. The 128-node array is configured in 4 petals to minimize rover travel on the lunar surface, with a deliberate asymmetry to improve imaging performance. The array is deployed on 4 successive lunar days, one petal of 32 nodes deployed in the 10–14 days of sunlight, with the rover hibernating during lunar night.

The FARSIDE mission architecture and concept of operations were developed in a sequence of studies conducted at the Jet Propulsion Laboratory (JPL) by JPL's concurrent design facility, Team X. The sequence of Team X studies assessed the technical feasibility of the instrument, the mission as a whole, and the cost realism for implementation as a Probe-class mission. The pre-reserve total cost for the mission is ~$1.1B, including the launch vehicle and the nuclear material power sources for the base station and the 128 receiver elements comprising the science instrument. The development cost is $800M, ~$110M for the instrument (payload) and ~$510M for the flight system, which consists of the base station and deployment rover integrated into a large lunar lander. The Blue Origins Blue Moon Lander was selected as a reference lander for the design. The total mission cost estimate for FARSIDE, after applying NASA- and JPL-standard cost reserves of 30% during development and 15% during operations is ~$1.3B.

FARSIDE is timely. Increasing interest in lunar exploration from NASA is driving the development of the Lunar Gateway and supporting technology, and stimulating the advancement of progressively more capable lunar landers. FARSIDE will take advantage of these investments from NASA's Artemis program, which are expected to reach sufficient maturity by the mid-2020's to support a mission in the 2025–2035 time frame. The FARSIDE instrument and the computing elements on the base station will need modest engineering development; no new technology is required.

## 1.1 Science Traceability Matrix

| Investigation | Goals | Objectives | Scientific Measurement Requirements: Physical Parameters | Scientific Measurement Requirements: Observables | Instrument Functional Requirements | Instrument Predicted Performance | Mission Functional Requirements Common to all Investigations | Mission Functional Requirements Specific to Each Investigation |
|---|---|---|---|---|---|---|---|---|
| Exoplanets and Space Weather | NASA Science Plan 2014<br>• Discover and study planets around other stars, and explore whether they could harbor life.<br>New Worlds, New Horizons (2010 Decadal Survey)<br>• Do habitable worlds exist around other stars, and can we identify the telltale signs of life on an exoplanet?<br>• Discovery area: Identification and characterization of nearby habitable exoplanets.<br>Exoplanet Science Strategy (National Academies of Sciences 2018) Goal 2: to learn enough about the properties of exoplanets to identify potentially habitable environments and their frequency, and connect these environments to the planetary systems in which they reside.<br>• The presence and strength of a global-scale magnetic field is a key ingredient for planetary habitability. | E1: Determine the prevalence and strength of large-scale magnetic fields on rocky planets orbiting M dwarfs and assess the role of planetary magnetospheres in the retention and composition of planetary atmospheres and planetary habitability. | Planetary magnetic field strength (proportional to frequency).<br><br>Local stellar wind velocity.<br><br>Planetary rotation period and assessment of the presence of a convective interior for a sample of rocky planets orbiting M dwarfs out to 10 pc. | Planetary radio flux: < 150 μJy (in the 150 kHz–250 kHz band).<br><br>Frequency range: 150 kHz–2 MHz band. | **Noise Equivalent Flux** (for 60 second integration, bandwidth = half central frequency): 93 mJy @ 200 kHz<br><br>**Spatial Resolution (FWHM):** 10 deg @ 200 kHz<br><br>**Spectral Resolution:** < 30 kHz<br><br>**Temporal Resolution:** < 60 seconds<br><br>**Minimum Frequency:** < 150 kHz<br><br>**Maximum Frequency:** > 2 MHz<br><br>**Number of Frequency Channels in band:** > 64<br><br>**Sky Coverage:** > 5,000 sq. degrees | **Noise Equivalent Flux:** 93 mJy @ 200 kHz<br><br>**Pointing Resolution (FWHM):** 10 deg @ 200 kHz<br><br>**Spectral Resolution:** 28.5 kHz<br><br>**Temporal Resolution:** 60 seconds<br><br>**Minimum Frequency:** 100 kHz<br><br>**Maximum Frequency:** 40 MHz<br><br>**Number of Frequency Channels in band:** 1400<br><br>**Sky Coverage:** 10,000 sq. degrees<br><br>**Noise Equivalent Brightness Temperature Sensitivity:** (@ 15 MHz over 500 kHz bandwidth): 5 mK<br><br>**Antenna Beam Size:** 10,000 sq. deg<br><br>**Antenna Beam Pattern Knowledge:** 50 dB | **Location:** Latitude and longitudes within 65 degrees of the anti-Earth point (required to suppress RFI from Earth by ~80dB). | **Observation time:** > 2400 hours |
| | | E2: Determine whether the largest stellar flares are accompanied by comparably large CMEs that can escape the corona of the star to impact the space environment of orbiting exoplanets.<br><br>E3: Determine the space weather environment of rocky planets orbiting M dwarfs during extreme space weather events and assess whether such events play a decisive role in atmospheric retention and planetary habitability.<br><br>E4: Determine the impact of extreme space weather events on exoplanets orbiting Solar type (FGK) stars and assess whether such events play a decisive role in atmospheric retention and planetary habitability. | Stellar radio bursts from particles accelerated in magnetic fields that vary in time and frequency (dynamic spectrum) depending on the local plasma frequency, as well as the velocity and mass of the associated CME shock (Type II) or electron beam (Type III). | Radio burst dynamic spectrum: sensitivity 93 mJy @ 200 kHz over 60 seconds.<br><br>Frequency range: 150 kHz–2 MHz band. | **Noise Equivalent Flux** (for 60 second integration, bandwidth = half central frequency): 93 mJy @ 200 kHz<br><br>**Spatial Resolution (FWHM):** 10 deg @ 200 kHz<br><br>**Spectral Resolution:** < 30 kHz<br><br>**Temporal Resolution:** < 60 seconds<br><br>**Minimum Frequency:** ≤ 150 kHz<br><br>**Maximum Frequency:** > 2 MHz<br><br>**Number of Frequency Channels in band:** > 64<br><br>**Sky Coverage:** > 5,000 sq. degrees | | | |
| Cosmology | "Explore how (the Universe) began and evolved"<br>NASA Science Plan (2014)<br><br>"What is the nature of dark matter?"<br>Astro2010<br><br>"Resolve the structure present during the dark ages and the reionization epoch"<br>NASA Astrophysics Roadmap | C1: Determine if excess cooling beyond adiabatic expansion in standard cosmology and exotic physics (e.g., baryon-dark matter interactions) are present in the Dark Ages with > 5σ confidence. | Redshift-dependent *mean* brightness temperature variation of the cosmic radio background at the level of ~100 mK due to the spin-flip transition of neutral hydrogen.<br><br>Redshift range approx. (50 < z < 130) | Brightness temperature: a ~40 mK absorption feature between 11–28 MHz against the cosmic radio background, globally averaged over > 10 deg^2.<br><br>Frequency range approx. 11–28 MHz (corresponding to 50 < z < 130).<br><br>Frequency resolution of <50 kHz to allow foreground & RFI mitigation: re-binned frequency resolution of <500 kHz to resolve the 21cm absorption feature and allow systematic checks.<br><br>Astrophysical foreground mitigation to better than 10^5 level in spectral domain. | **Noise Equivalent Brightness Temperature Sensitivity** (@ 15 MHz over 500 kHz bandwidth): < 8 mK<br><br>**Antenna Beam Size:** field-of-view > 10 deg^2<br><br>**Minimum Frequency:** < 10 MHz<br><br>**Maximum Frequency:** > 30 MHz<br><br>**Spectral Resolution:** < 50 kHz<br><br>**Antenna Beam Pattern Knowledge:** To a level of < 50 dB. | | | **Observation time:** > 1800 hours |

# 2 SCIENCE

## 2.1 The Magnetospheres and Space Weather Environments of Habitable Planets

The discovery of life on a planet outside our solar system is at the heart of NASA's Science Mission Directorate. Such a discovery may arrive within the next few decades and is the focus of a number of planned and concept NASA missions. The most likely avenue involves spectral observations of biosignature gases, such as $O_2$, $O_3$, $CH_4$, and $CO_2$ on an Earth-like planet orbiting a nearby star. A tiered approach will involve discovery (e.g., *TESS*, ground-based radial velocity (RV) surveys), characterization and eventual deep coronagraph-assisted observations by missions such as *JWST, WFIRST, HabEx* and *LUVOIR* as well as ground-based extremely large telescopes.

***The Active Young Sun:*** However, it is becoming increasingly apparent that both the selection of candidate exoplanets for deep searches for biosignatures, and interpretation of the observed atmospheric composition, must take into account the space weather environment of the host star, and whether the planet possesses a large-scale magnetosphere capable of retaining an atmosphere within this space environment. Studies of young solar analogs have shown that the enhanced magnetic activity of the young zero-age main sequence Sun [Ribas et al. 2005], powered by its rapid rotation, was a major factor in defining the atmospheric properties of the solar system planets. The enhanced radiative output at higher energies during flares leads to strong expansion of planetary atmospheres [Lammer et al. 2003; Kulikov et al. 2007; Grießmeier al. 2005]. Simultaneously, the stellar wind of a young star is much denser and faster, compressing the magnetosphere, particularly during coronal mass ejections (CMEs), which are also presumed to be much more frequent [Grießmeier et al. 2015][1], as well as associated solar energetic particle (SEP) events. Combined, these effects would lead to an exosphere which is much closer to the magnetosphere than the present-day Earth, and strong atmospheric erosion in the absence of a strong planetary magnetic field [Khodachenko et al. 2007; Lammer et al. 2007]. Atmospheric model calculations by Kulikov et al. [2007] investigated this effect in the case of the Earth, Venus and Mars in the early solar system and conclude that the lack of water in the atmosphere of Venus could be attributed to the absence of a strong magnetic field. The presence of a much stronger planetary magnetosphere likely prevented a similar process occurring on Earth. The long-term impact of such activity was recently established in dramatic fashion by the Mars Atmosphere and Volatile EvolutioN (*MAVEN*) mission, which confirmed that ion loss due to solar CMEs early in Mars history likely severely depleted its atmosphere [Jakosky et al. 2015].

***Extreme M Dwarfs:*** The impact of magnetic activity on planets orbiting M dwarfs in particular has become a topic of increasing significance for stars in the mass range that can host evolved exoplanets (age > 1 Gyr), 95% are M dwarfs. Kepler has shown that most M dwarfs harbor small planets, with $2.5\pm0.2$ planets per M dwarf with radii 1–4 Earth radii and periods shorter than 200 days [Dressing and Charbonneau 2013, 2015]. Together, this implies that the nearest potentially habitable planet orbits an M dwarf, and indeed statistics from Kepler suggest such a system is < 3 pc from Earth [Dressing and Charbonneau 2015]. Recent spectacular examples include the Trappist-1 system [Gillon et al. 2017] and the detection of a possible terrestrial planet in the habitable zone of the nearest star, Proxima Centauri [Anglada-Escude et al. 2016].

However, many M dwarfs are known to be particularly magnetically active, flaring frequently and with much higher energy than produced in solar flares [Osten et al. 2010]. Moreover, M dwarfs have longer spin-down timescales (> 1 Gyr) and are thus magnetically active for a much longer period of time than G dwarfs like our Sun [West et al. 2008]. Studies of possible flares and CME events on planets in the habitable zone around such stars suggest that these events severely impact the ability of such planets to retain their atmospheres [Khodachenko et al. 2007; Lammer et al. 2007]. The habitable zone—as defined by Kasting et al. [1993] around M dwarfs is much closer to the parent star than the solar case ($\leq 0.2$ AU). As well as resulting in exposure to X-ray and extreme ultraviolet (EUV) flux 10–100 times higher than the present solar flux at 1 AU [Scalo et al. 2007], these tighter orbits result in tidal locking timescales < 100 Myr for terrestrial planets [Grießmeier et al. 2005], resulting in much reduced magnetic moments. Planetary magnetic fields act as shields against the bombardment of energetic particles in stellar winds, the absence of which can lead to catastrophic loss of atmosphere due to CME-induced ion pick up, particularly in $CO_2$ rich atmospheres [Lammer et al. 2007]. The latter authors show that the expected atmospheric loss may be severe even in the case of a magnetic moment equivalent to Earth's, such that a planet may require a significantly stronger magnetosphere than Earth to sustain an Earth-like atmosphere and biosphere in the space weather environment of an M dwarf.

---

[1] And all references therein.

***Detecting Stellar CMEs and SEP events:*** While these modeling results paint a potentially bleak picture, there is a very large degree of uncertainty in the space environments to which exoplanets are exposed. Other than the Sun, no main sequence star has been detected to produce a CME or an SEP event. Similarly, detection of exoplanet magnetic fields has yet to be achieved and remains the most crucial ingredient in assessing planetary habitability in the context of stellar activity.

Solar CMEs and SEP events can be accompanied by radio bursts at low frequencies, particularly so-called Type II bursts, as well as a subset of Type III bursts (complex Type III-L bursts). The emission is produced at the fundamental and first harmonic of the plasma frequency and provides a diagnostic of the density and velocity (few 100 to >1000 km/s) near the shock front, while the flux density of the burst depends sensitively on the properties of the shock and solar wind [Cairns et al. 2003]. Ground-based radio astronomy can only trace such events to a heliocentric distance of a few solar radii, whereas space-based radio antennas can trace the propagation of shocks out to the Earth and beyond, which is particularly relevant for characterizing geoeffective CMEs and SEP events. The detection of equivalent interplanetary Type II and III events from stars other than the Sun is one of the goals of the FARSIDE array.

Solar radio bursts are likely the most intense sources of extraterrestrial radio emission ever observed from Earth (Figure 2.1-1), reaching a flux density > $10^{-14}$ W m$^{-2}$ Hz$^{-1}$ (>$10^{12}$ Jy). However, the brightest bursts are rare and peak at frequencies <10 MHz [Saint-Hilaire et al. 2013], which can account for the non-detection of such bursts from solar-type stars. The non-detection of such events from active M dwarfs has been surprising [Villadsen and Hallinan 2019; Crosley and Osten 2018]. Even slow rotating M dwarfs produce a "superflare" ($10^{34}$ erg) each month, with >100x larger energy than any recorded on the Sun [Yang et al. 2017], and a correlation is observed between the flare energy and mass (or kinetic energy) of CMEs observed on the Sun [Aarnio et al. 2011; Osten and Wolk 2015] which, if applicable to M dwarfs, should produce very large CMEs, and associated luminous Type II bursts [Cairns et al. 2003].

However, this scaling law may not extend up to the flare energies observed for M dwarfs. For example, it is possible that CMEs are confined or significantly suppressed by the presence of a strong large-scale magnetic field [Alvarado-Gómez et al. 2018]. Although the largest events would still escape to interplanetary distances, this would lessen the potential impact of CMEs/SEPs on the atmospheres of orbiting exoplanets. An alternative possibility is that CMEs are present, but are not detected, simply because the Alfvén speed is too high in the coronae of M dwarfs for a shock to form [Villadsen and Hallinan 2019; Mullan and Paudel 2019]. The presence of a shock is an observed necessary condition for Type II emission and a prerequisite for the generation of certain classes of SEP events. Shocks can potentially form at much greater distances from the star, where the Alfvén speed drops below the velocity of the CME, but the associated Type II burst would then be at frequencies below 10 MHz, and undetectable to ground-based radio telescopes.

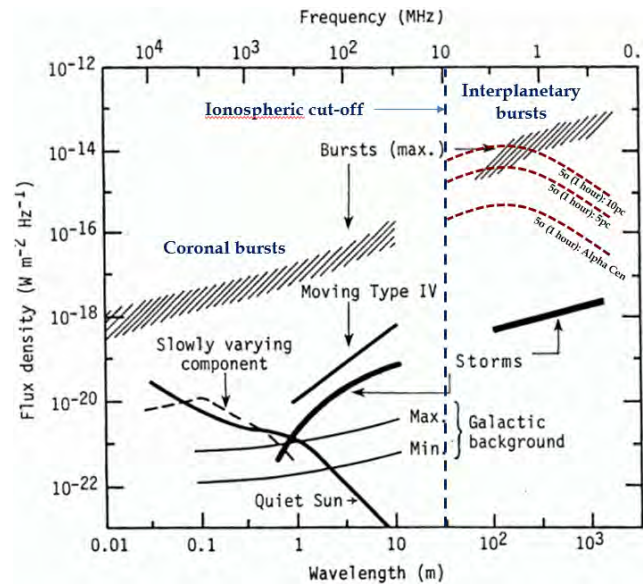

**Figure 2.1-1.** The flux density and frequency/wavelength of the brightest radio sources observed from Earth orbit. The brightest phenomena are solar radio bursts at all frequencies, particularly Type II and III bursts [Gopalswamy 2004]. Interplanetary bursts are the most luminous and are not detectable from the ground. FARSIDE can detect such events out to >10 pc.

FARSIDE would detect the equivalent of the brightest Type II and Type III bursts out to 10 pc at frequencies below a few MHz. By imaging >10,000 square degrees every 60 seconds, it would monitor a sample of solar-type stars simultaneously, searching for large CMEs and associated SEP events. For the Alpha Cen system, with two solar-type stars and a late M dwarf, it would probe down to the equivalent of $10^{-15}$ W m$^{-2}$ Hz$^{-1}$ at 1 AU, a luminosity at which solar radio bursts are frequently detected at the lowest frequencies available to FARSIDE [Krupar and Szabo 2018]. The nearby young active solar-type star, Epsilon Eridani (spectral type K2), is another priority target. For the case of M dwarfs, FARSIDE would be able to detect Type II bursts formed at the distance where super-Alfvénic shocks should be possible for M dwarfs and directly investigate whether the relationship observed between solar flares and CMEs extends to M dwarfs. If it does, FARSIDE should also detect a very high rate of radio bursts from this population (Figure 2.1-2).

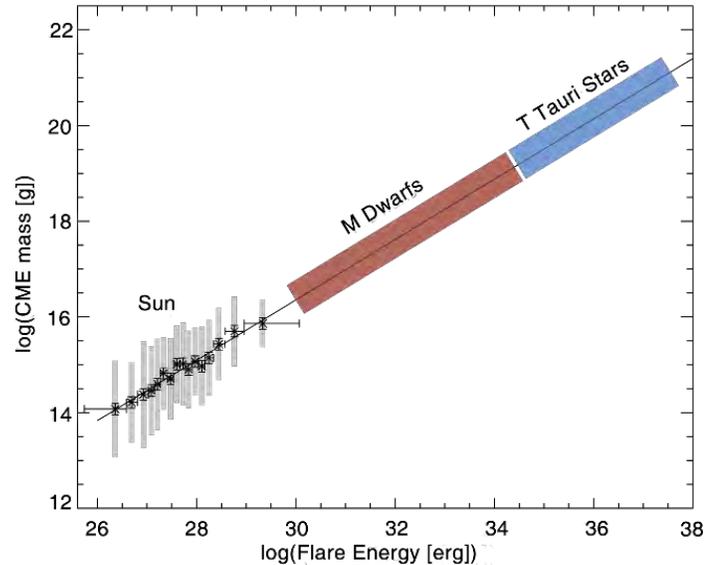

**Figure 2.1-2.** The relationship between flare energy and CME mass observed for the Sun [Aarnio et al. 2011], extrapolated to flare energies observed on M dwarfs [Yang et al. 2017] and young stars.

***Exoplanet Magnetospheres:*** All the magnetized planets in our solar system, including Earth, produce bright coherent radio emission at low frequencies, predominantly originating in high magnetic latitudes and powered by auroral processes [Zarka 1998]. Detection of similar radio emission from candidate habitable planets is the only known method to directly detect the presence and strength of their magnetospheres. The temporal variability of this radio emission can also be used to determine the rotation periods and orbital inclinations of these planets. The radio emission, generated by the electron cyclotron maser instability, is produced at the electron cyclotron frequency, ~2.8 × $B$ MHz, where $B$ is the magnetic field strength in Gauss at the source of the emission. Notably, only Jupiter has a strong enough magnetic field to be detected from the ground. The radio emission of Earth, Saturn, Uranus and Neptune are predominantly confined to frequencies < 2 MHz requiring a space-based instrument for detection [Zarka 1998].

Extending to the exoplanet domain requires a very large collecting area at low frequencies, with the first detections likely to be ground-based. Indeed, in a recent breakthrough, radio emission has been detected from a possible free-floating planetary mass object (12.7 ±1 M$_{jup}$; [Kao et al. 2018]). This is the first detection of its kind and confirmed a magnetic field much higher than expected, >200 times stronger than Jupiter's, reinforcing the need for empirical data. However, the detection of the magnetic fields of candidate habitable planets will almost certainly require a space-based array, if the magnetic fields are within an order of magnitude in strength of Earth's magnetic field. Detection of the magnetic field of planets orbiting in the habitable zone of M dwarfs is particularly key, as such planets may require a significantly stronger magnetosphere than Earth to sustain an atmosphere [Lammer et al. 2007].

The detection of radio emission from planets orbiting nearby stars is very sensitive to the stellar wind conditions imposed by its parent star and can serve as an indirect diagnostic of its velocity and density. It is possible to estimate the expected radio power from such planets, based on scaling laws known to apply to radio emission from the solar system planets. These scaling relations are not only descriptive but predictive, with the luminosity of both Uranus and Neptune predicted before the Voyager 2 encounters and found to be in excellent agreement with the measurements [Desch and Kaiser 1984; Desch 1988]. Extrapolations to terrestrial exoplanets in the habitable zone of M dwarfs, and thus embedded within a dense stellar wind environment, predict radio luminosities that are orders of magnitude higher than the Earth [Vidotto et al. 2019; Grießmeier 2015]. During enhanced solar wind conditions, the Earth's radio emission can increase in luminosity by orders of magnitude (Figure 2.1-3) and a similar effect is expected for exoplanets. Therefore, as is the case for detecting Type II bursts associated with CMEs, it is essential to have the capability to monitor large numbers of planets simultaneously to detect periods where exoplanets greatly increase in luminosity.

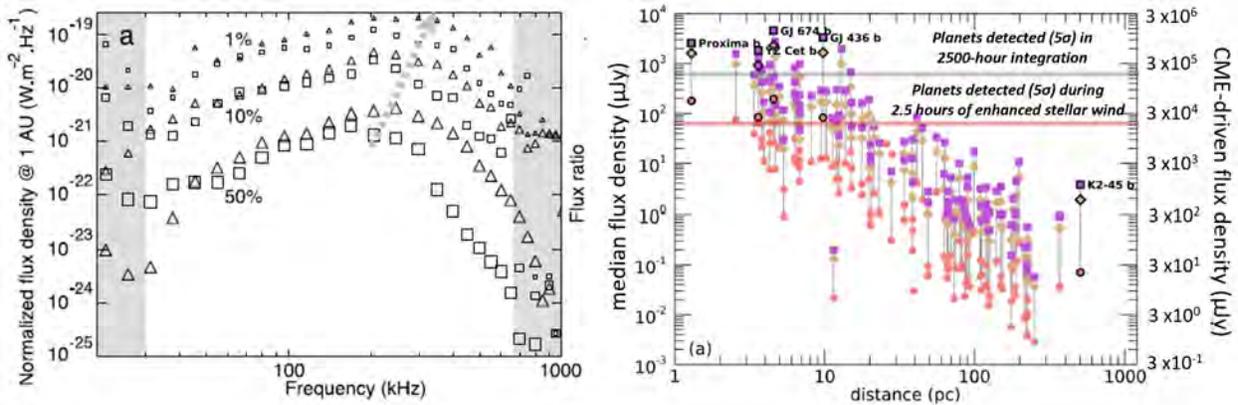

**Figure 2.1-3.** *Left*: The radio spectrum of Earth's auroral kilometric radiation for 50%, 10% and 1% of the time. The flux density across the entire detectable band can vary by factors of a few hundred [Lamy et al. 2010]. *Right*: The predicted average flux radio density at ~280 kHz of known exoplanets orbiting M dwarfs from [5], assuming a magnetic field of 0.1 G (10% of Earth's magnetic field). Purple, yellow and red data points reflect different models for the quiescent stellar wind. The horizontal grey line represents the 5σ detection limit (y-axis on left-hand side) on the median flux density for the auroral radio emission detectable from FARSIDE in a 2500-hour integration. The horizontal red line represents the 5σ detection limit (y-axis on right-hand side) in a 2.5-hour integration during enhanced stellar wind conditions, e.g., during a CME, assuming a similar degree of variability as observed for Earth's auroral kilometric radiation. A large sample of planets, including some nearby candidate habitable planets (such as Proxima b), are detectable.

FARSIDE will image 10,000 deg$^2$ of sky visible from the lunar far side every 60 seconds, in 1400 channels spanning 150 kHz–40 MHz. The sensitivity at the lower end of the band, enabled by the unique environment of the lunar far side, is particularly well suited to the search for radio exoplanets. In each FARSIDE image, there will be ~2,000 stellar/planetary systems within 25 pc. At these low frequencies, the spatial resolution is limited to a few degrees due to scattering in the interstellar medium and interplanetary medium. However, with 1 stellar system within 25 pc per ~5 square degrees, identification of the source of detected emission should be straightforward, particularly as FARSIDE is most likely to detect exoplanets within 10 pc (Figure 2.1-3).An advantage at these low frequencies, is that the galaxy is optically thick, and thus long integrations are possible without classical confusion noise. Over the course of a 2-year observing program, FARSIDE will have accumulated an average of 4,000 hours of observing time on each of 8,000 stellar planetary systems within 25 pc. Radio exoplanets will be identified by two means:

1. Deep 2,500 hour integrations will be used to place limits on the average radio power from radio exoplanets.

2. Individual 2.5-hour integrations will be searched for heightened emission, expected during CME interaction.

Figure 2.1-3 highlights that FARSIDE would detect the radio emission from a population of Earth-sized, and super-Earth sized planets orbiting nearby M dwarfs, including a number of candidate habitable

planets, providing the first measurements of terrestrial planet magnetospheres outside our solar system. The radio emission can be distinguished from that of the star by rotational modulation, as well as circular polarization which is largely absent for interplanetary radio bursts. Detection of magnetospheres, if present, would identify the most promising targets for follow-up searches for biosignatures, as well as providing a framework for comparative analysis of exoplanet magnetospheres and atmospheric composition.

An intriguing additional possibility involves using FARSIDE to identify optimal windows in which the possibility of detecting biosignature gases will be greatly enhanced. It has been demonstrated, using Proxima b as a test case, that the power in the O I 5577Å auroral line from an Earth-like atmosphere will be orders of magnitude larger for an exoplanet orbiting an M dwarf [Luger et al. 2017], with even greater enhancement during periods of strong magnetospheric disturbance (auroral power ~1 TW–10 TW). During the latter such periods, the planet-star contrast ratio in this line may be as low as $10^{-4}$–$10^{-5}$, with detection possible in < 1 day for space-based coronagraphic telescopes such as HabEx. During such periods, the auroral radio emission would be greatly enhanced. Monitoring of confirmed radio emitting exoplanets by FARSIDE can therefore be used to identify windows where the possibility of detection of an oxygen-rich atmosphere is greatly enhanced, triggering deep exposures at optical wavelengths.

## 2.2    Dark Ages Hydrogen Cosmology

FARSIDE enables precision measurements of the Dark Ages 21-cm signal at frequencies 10–40 MHz corresponding to redshifts z≈140-35. Such observations have enormous potential as a powerful new test of the standard ΛCDM cosmological model in the early Universe as they would provide constraints on any exotic physics operating at early times, including dark matter interactions and non-thermal effects in the low-frequency tail of the cosmic microwave background (CMB) radiation [Burns et al. 2019].

Figure 2.2-1 places the Dark Ages and Cosmic Dawn epochs into perspective. After the Big Bang, the Universe was hot, dense, and nearly homogeneous. As the Universe expanded, the material cooled, condensing after ~400,000 years (z~1100) into neutral atoms, freeing the CMB photons. The baryonic content during this pre-stellar Dark Ages of the Universe consisted primarily of neutral hydrogen. About fifty million years later, gravity propelled the formation of the first luminous objects – stars, black holes, and galaxies – which ended the Dark Ages and commenced the Cosmic Dawn—see e.g., Loeb and Furlanetto [2013]. These first stars (Pop III) likely differed dramatically from stars we see nearby, as they formed in vastly different environments—see e.g., Abel et al. [2002].

This transformative event marked the first emergence of structural complexity in our Universe, but no currently planned mission can make observations this far back in time (z≳35, or ≲80 Myrs after the Big Bang). While JWST, WFIRST, and a suite of ground-based facilities will observe the Universe as it was ≳ 300 Myrs after the Big Bang (z ≲ 15, and especially focus on the Reionization era when distant galaxies ionized the gas between them, up to about a billion years after the Big Bang), none now contemplate observing the true first stars and black holes—see e.g., Behroozi and Silk [2015]—much less the Dark Ages that preceded them. For example, CMB observations of Thomson scattering measure the integrated column density of ionized hydrogen, but only roughly constrain the evolution of the intergalactic medium—e.g., Mesinger et al. [2012]. Ly-$\alpha$ absorption from QSOs only confines the end of reionization at relatively late times, z~7 or 770 Myrs after the Big Bang—see e.g., McGreer et al. [2014]. Observations with HST and JWST will simply find the brightest galaxies at high redshifts (z≲15), and thus any inferences drawn about the high-z galaxy population as a whole depend upon highly uncertain assumptions about the faint-end slope of the luminosity function—see e.g., Bouwens et al. [2015]. The Hydrogen Epoch of Reionization (EoR) Array [HERA; DeBoer et al. 2017] will observe the neutral hydrogen (HI) power spectrum from ~50–200 MHz (z = 27 − 6, 117 − 942 Myrs) probing the Cosmic Dawn and EoR epochs but not the Dark Ages—see e.g., Ewall-Wice et al. [2016].

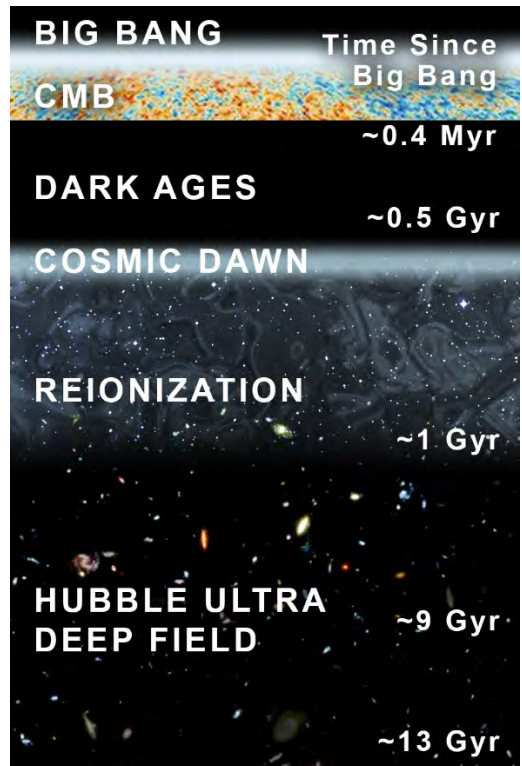

**Figure 2.2-1.** The pre-stellar (Dark Ages), first stars (Cosmic Dawn), and Reionization epochs of the Universe can be uniquely probed using the redshifted 21-cm signal. This history is accessible via the neutral hydrogen spin-flip background [Loeb and Furlanetto 2013]. FARSIDE would observe redshifts $z \approx 140 - 35$ ($v = 10 - 40$ MHz), corresponding to 10–80 Myrs after the Big Bang.

FARSIDE will employ two complementary approaches to observe the Dark Ages (z ~ 140 − 35, ~ 10 − 80 Myrs after the Big Bang) using the highly redshifted 21-cm signal. First, the sky-averaged Global signal or monopole can be observed by a spectrometer connected to individual dipole antennas operating in the total power mode. This is analogous to the CMB blackbody spectrum that was precisely measured by COBE. But, unlike the CMB, the spectral 21-cm line enables the line of sight evolution of the intergalactic medium (IGM) to be measured (Figure 2.2-2). The observed brightness temperature, $T_b(\nu)$, is a gauge of the evolution of the neutral hydrogen fraction in the IGM, along with the radio background and gas temperature (see Eq. 2.2-1). In the Dark Ages before any luminous sources turn-on, these physical parameters depend only upon the cosmic adiabatic expansion of the Universe. Any deviation from the standard cosmological model requires novel additional physics. Therefore, a measurement of the hydrogen Global signal is not only a powerful test of the standard cosmological model in an epoch heretofore unobserved but also a flag of exotic physics in the early Universe.

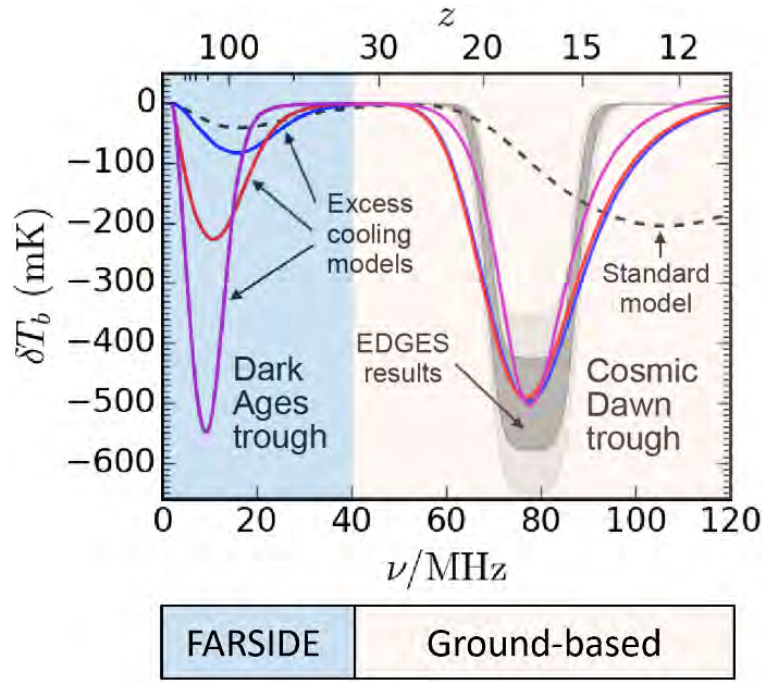

**Figure 2.2-2. The Dark Ages 21-cm absorption trough is a sensitive probe of cosmology.** The black dashed curve shows the brightness temperature (relative to the radio background) in a standard cosmological model with adiabatic cooling. The shape at z > 30 is independent of astrophysical sources. The grey contours show the 1 & 2-sigma absorption bands inferred by EDGES. The solid color curves are parametric models that invoke extra cooling to match the amplitude of the EDGES signal in the Cosmic Dawn frequency range but can also dramatically affect the Dark Ages absorption trough at z > 50. FARSIDE will cleanly distinguish between the standard cosmology model with adiabatic expansion and models with added cooling, possibly due to new interactions with dark matter, at >5-sigma level.

Second, spatial fluctuations in the 21-cm Dark Ages signal are governed almost entirely by well-understood linear structure formation, the same physics used to interpret *Planck* and other observations of the CMB power spectrum, allowing precise predictions of the expected signal within the standard cosmological model. Interferometric measurements of fluctuations in the 21-cm radio background can therefore quantify departures from these well-constrained predictions for structure formation and provide important new insights into the nature of dark matter. Such observations could also measure the ultimate number of linear modes of the Universe and exquisitely constrain the non-Gaussianity of structures to test the inflationary paradigm.

*Redshifted 21-cm Global Spectrum:*

The most promising method to measure the properties of the Dark Ages and the Cosmic Dawn in the near term is the highly redshifted Global or all-sky 21-cm spectrum illustrated in Figure 2.2-2—see e.g., Pritchard and Loeb [2012]. The "spin-flip" transition of neutral hydrogen emits a photon with a rest frame wavelength of 21-cm ($\nu = 1420$ MHz). As neutral hydrogen composed the majority of baryonic matter in the early Universe, its emission in the Global 21-cm spectrum provides crucial cosmological information from the Dark Ages through the Cosmic Dawn, as well as the EoR. Its signal is generated by a weak hyperfine transition, but its ubiquity means it can be detected even with a simple radio dipole experiment.

The curves in Figure 2.2-2 show broad spectral features that are common to virtually all 21-cm models of the early Universe. Its diagnostic trait, brightness temperature, is driven by the evolution of the ionization fraction ($x_{HI}$) and spin temperature $T_S$ (measure of the fraction of atoms in the two spin states) of HI relative to the radio background temperature $T_R$—see e.g., Furlanetto et al. [2006]; Shaver et al. [1999] as given by:

$$\delta T_b \simeq 27\ \bar{x}_{H\,I}(1+\delta)\left(\frac{\Omega_{b,0}h^2}{0.023}\right)\left(\frac{0.15}{\Omega_{m,0}h^2}\frac{1+z}{10}\right)^{1/2}\left(1-\frac{T_R}{T_S}\right)\ mK. \qquad \text{(Eq. 2.2-1)}$$

Here, $\delta$ is the baryon overdensity relative to the mean density of the Universe (~0 for the all-sky signal), h is the normalized Hubble constant, and $\Omega_b$ ($\Omega_m$) is the baryon (total mass) abundance parameter. Because of cosmic expansion, the observed frequency $\nu_0$ corresponds to a redshift z through the relation $\nu/\nu_0 = 1+z$, where $\nu = 1420$ MHz is the rest frame frequency. The greater the redshift of the hydrogen, the older it is, so $\delta T_b(\nu)$ probes the growth of structure in the early Universe.

In the standard $\Lambda$CDM cosmological model, the lowest frequency (corresponding to the highest redshift and earliest time) spectral absorption feature in each curve of Figure 2.2-2 is called the "Dark Ages trough" ($\nu < 40$ MHz). It is purely cosmological and thus relatively simple to interpret because there are no stars or galaxies at this epoch to complicate the signal. At $z \gtrsim 30$ ($\lesssim 100$ Myrs after the Big Bang), cosmic expansion drives a decoupling between the spin temperature and the radio background radiation temperature ($T_R > T_S$) producing a broad absorption feature in the 21-cm spectrum. The standard cosmological model (dashed black curve in Figure 2.2-2) makes a precise prediction for the central frequency ($\approx 18$ MHz) and brightness temperature ($\approx 40$ mK) for this feature. Any departure from this model would indicate the need for additional physics such as, e.g., non-gravitational interactions between baryons and dark matter—see e.g., Barkana [2018]; Muñoz and Loeb [2018]; Berlin et al. [2018]; Muñoz et al. [2018], as shown by the color parametric curves in Figure 2.2-2. Thus, the low frequency 21-cm spectrum offers a unique and robust probe of the standard cosmological model.

Star formation, which probably begins around z~30, produces Ly-$\alpha$ photons that couple the gas temperature to the hyperfine spin-flip transition via the Wouthuysen-Field mechanism [Pritchard and Loeb 2012]. As the gas adiabatically cooled faster than the CMB radiation, this results in a broad absorption feature in the 21-cm spectrum, the "Cosmic Dawn trough", starting at $\approx 50$ MHz (Figure 2.2-2). Thus, this extremum point where the spectrum turns downward is a measure of when the first stars "turn on" and the gradient of the spectrum provides constraints on the nature of the first stars (e.g., the ratio of first generation, Pop III, to second generation, Pop II, stars; [Burns et al. 2017]. The massive black hole remnants of these first stars accrete gas and generate X-ray emission, which heats intergalactic neutral hydrogen and drives the 21-cm signal back towards the CMB temperature (see troughs at z~23-13, $\nu$~60-100 MHz in Figure 2.2-2).

Recent results [17, 18] from the *Experiment to Detect the Global EoR Signature* (EDGES) suggest the presence of a strong feature in the 21-cm spectrum at $\approx$78 MHz ($z \approx$17), within the time period expected for the Cosmic Dawn trough. Figure 2.2-2 shows examples of deviations from the standard cosmology for phenomenological models of added cooling to the primordial neutral hydrogen [Mirocha and Furlanetto 2019] that could be produced by, e.g., previously unanticipated interactions between baryons and dark matter particles. The black dashed curve assumes a standard $\Lambda$CDM cosmology model and that the earliest galaxies evolve according to smooth extrapolations of the known high-z galaxy population [Burns et al. 2017; Mirocha et al. 2016]. While the difference in the observed EDGES redshift can be explained within the standard model, the drop in brightness is about 3 times greater than allowed by adiabatic gas cooling at the frequency corresponding the trough minimum relative to the standard model (Figure 2.2-2). The Dark Ages trough, at $\nu < 30$ MHz and inaccessible from the ground due to ionospheric effects, is produced purely by cosmology and is thus cleaner because there are no stars to complicate the signal. This trough is a prime target for FARSIDE. If the measured redshifted 21-cm signal differs from that of the standard cosmological model, new physics is required.

There are three possible explanations for the deep Cosmic Dawn trough (grey bands) in Figure 2.2-2. First, it might be explained by an increase in the radio background above the CMB ($T_R$ in Eq. 2.2-1; e.g., Feng and Holder [2018]; Fialkov and Barkana [2019]), sourced by synchrotron emission in the first star-forming galaxies or Active Galactic Nuclei—see e.g., Mirocha and Furlanetto [2019]; Ewall-Wice et al. [2018]—or heated by neutrinos [Chianese et al., 2019], dark matter annihilation [Fraser et al. 2018], or even superconducting strings [Brandenberger et al. 2019]. Second, the trough could also be explained by a change in the cosmological parameters (e.g., baryonic content or Hubble constant as indicated by $\Omega_{b,0}h^2$ in Eq. 2.2-1). However, this seems unlikely given current observational constraints. A third possibility, which has received much attention, is that the trough could be produced through enhanced cooling of the hydrogen ($T_S$ in Eq. 2.2-1) via, e.g., Rutherford scattering of baryons off of dark matter—see e.g., Muñoz et al. [2015]; Barkana [2018]; Fialkov et al. [2018]. However, independent constraints suggest that this source of scattering could not constitute all of the dark matter in the Universe but rather a sub-percent fraction [Muñoz and Loeb 2018; Berlin et al. 2018; Kovetz et al. 2018]. Identifying excess cooling in the early Universe could thus provide the first evidence that there is more than one kind of dark matter. Indeed, the timing of the signal alone can constrain the properties of any warm component of dark matter—see e.g., Safarzadeh et al. [2018]; Schneider [2018]; Lidz and Hui [2018].

In Figure 2.2-2, we show three examples of "excess cooling" models (solid color curves) that, while consistent with the EDGES trough at 78 MHz, make different predictions for the Dark Ages signal. Based upon our parametric models, these curves demonstrate the effects of both different cooling rates and time of the cooling, while adjusting astrophysical parameters in order to preserve the 78 MHz feature seen by EDGES. The 78 MHz Cosmic Dawn trough, even though suggestive of exotic physics such as dark matter interactions, describes a complicated epoch in which multifaceted astrophysics such as star formation, ionization, and black hole X-ray heating occur. The Dark Ages absorption feature, to be measured by FARSIDE, reflects the simple state of the Universe before the formation of the first stars. FARSIDE observations have the potential to resolve these ambiguities and cleanly constrain the origin and characteristics of any source of additional cooling.

***FARSIDE Dark Ages Global Measurement:*** The major challenge for Global signal observations is the presence of bright foregrounds that are $10^{4\text{-}6}$ times that of the 21-cm signal, which together with chromaticity of the antenna beam, can introduce added frequency structure into the observed foreground brightness temperature spectrum. FARSIDE mitigates these effects in several important ways. First, the array would provide high resolution maps of the sky foreground at multiple relevant frequencies. This facilitates the extraction of the 21-cm signal using a data analysis pipeline that we have built based on pattern recognition algorithms informed by training sets [Tauscher et al. 2018] that can be obtained from sky foreground observations, lab instrument measurements, beam simulations and 21-cm signal theory. Second, an orbiting calibration beacon would permit us to map the antenna beam in the far-field for the first time enabling correction for beam chromaticity effects.

The system temperature ($T_{sys}$) at these frequencies is driven by emission from the astrophysical foreground whose thermal noise is determined via the radiometer equation:

$$\sigma_{RMS} = T_{sys}(N\Delta\nu \cdot t)^{-1/2}, \qquad \text{(Eq. 2.2-2)}$$

where $\sigma_{RMS}$ is the RMS thermal noise, N is the number of antennas (used in auto-correlation mode), $\Delta\nu$ is the frequency bandwidth, and $t$ is the integration time. In our observation band, the non-thermal brightness temperature of the Galaxy increases with decreasing frequency ($T_{Galaxy} \sim 5000K[\frac{\nu}{50MHz}]^{-2.5}$). Thus, the integration time increases quickly as we observe at lower frequencies. Combining all N=128 auto-correlations of the array, at 15 MHz approximately 1800 hours of integration time is required to achieve an RMS noise level of ≤5 mK, assuming a frequency channel width of 0.5 MHz. FARSIDE would thus be able to distinguish the standard cosmology from added cooling models at better than $5\sigma$ significance.

### The Dark Ages 21-cm Power Spectrum:

In many ways, spatial fluctuations in the 21 cm absorption during the cosmic Dark Ages provide the ultimate cosmological observable. The simplest way to quantify these fluctuations is with the power spectrum, which characterizes the amplitude of the variations as a function of spatial scale. During this time, the 21 cm line traces the cosmic density field with most modes in the linear or mildly non-linear regime, allowing a straightforward interpretation of the measurement in terms of the fundamental parameters of our Universe [Lewis and Challinor 2007]. The lack of luminous astrophysical sources makes the Dark Ages signal a clean and powerful cosmological probe and also renders the 21-cm line the *only* observable signal from this era. Furthermore, the 21-cm line can be used to reconstruct a 3D volume (as compared to the 2D surface of the CMB) and is not affected by "Silk damping" on the smallest scales, which blurs out fluctuations in the CMB—meaning the number of accessible modes is enormous and is effectively only limited by the collecting area of the instrument. Because 21-cm measurements can probe these small scales inaccessible by other means, they enable stringent constraints on key probes of inflation, including the running of the matter power spectrum spectral index [Mao et al. 2008] and non-Gaussianity [Munoz et al. 2015b], as well as constraints on the curvature of the Universe and the mass of the neutrino [Mao et al. 2008].

The 21-cm power spectrum is also an exquisite probe of physics occurring during the Dark Ages themselves. All the physical processes that affect the Global signal described above, including the exotic scenarios proposed for explaining the EDGES signal [Barkana 2018; Muñoz and Loeb 2018; Berlin et al. 2018; Muñoz et al. 2018 and Mirocha et al., 2016; Feng and Holder 2018; Fialkov et al. 2018a; Ewall-Wice et al. 2018; Fraser et al. 2018; Muñoz et al. 2015; Fialkov et al. 2018b; Kovetz et al. 2018; Safarzadeh et al. 2018; Schneider 2018; Lidz and Hui 2018], also affect the power spectrum. But the latter has far more information so allows more precise tests of the scenarios. Moreover, many exotic processes imprint distinct signatures in the Dark Ages power spectrum (Figure 2.2-3) [Muñoz et al. 2018]. Even if the EDGES measurement is not confirmed, dark matter itself remains a mystery and the cosmic Dark Ages offer the cleanest astrophysical probe of dark matter physics on cosmic scales. For example, any warm dark matter will suppress the formation of small structures and hence the amplitude of the power spectrum on those scales.

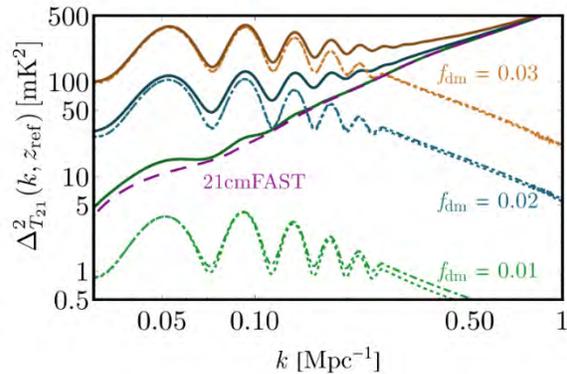

**Figure 2.2-3.** The 21-cm power spectrum can distinguish between different exotic physics scenarios during the Dark Ages. In these models, a fraction $f_{dm}$ of the dark matter is assumed to have a small charge; the oscillations in the power spectrum arise from the large-scale streaming of baryons relative to dark matter. The solid curves are the total power, while the others show the contribution from dark matter-baryon scattering. These models are compared to the standard cosmological model (labeled "21cmFAST"). Figure from [Muñoz et al. 2018].

Constraining such scenarios with observations targeting later epochs is complicated by the slew of baryonic feedback processes that also operate on small scales. As a result, 21-cm observations targeting epochs after the Cosmic Dawn suffer a degeneracy between astrophysical processes that halt star formation in low-mass dark matter halos and cosmological processes that suppress the formation of those dark matter halos. Observations aimed at the Dark Ages avoid this important degeneracy.

***FARSIDE Dark Ages Power Spectrum Pathfinder:*** There are many challenges to detecting the Dark Ages power spectrum. Intrinsically, it is an extremely faint signature, and sky-noise dominates any measurement. The signal strength predicted by standard cosmological theory will require a collecting area of order 5 square kilometers to be deployed above the Earth's ionosphere—far larger than the proposed FARSIDE array (see §3). Foreground emission complicates the picture even further. If residual foreground emission renders some modes unusable for cosmology, even more collecting area will be required to obtain a high significance 21-cm signal detection. FARSIDE is therefore an invaluable pathfinder for a larger 21 cm power spectrum instrument. It can both measure the intrinsic spectrum of the low-frequency foregrounds without the effects of the Earth's ionosphere and can be used as a testbed for foreground removal techniques. Such measurements will be necessary for determining the array size ultimately necessary for a 21-cm Dark Ages power spectrum measurement.

## 2.3 Additional Science

### 2.3.1 *Heliophysics*

During the solar illumination of lunar day, FARSIDE will observe solar radio bursts, which are one of the primary remote signatures of electron acceleration in the inner heliosphere. Our focus is primarily on two emission processes, Type II and Type III radio bursts (Figure 2.3-1). Type II bursts originate from suprathermal electrons ($E > 100$ eV) produced at shocks. These shocks are generally produced by CMEs as they expand into the heliosphere with Mach numbers greater than unity. Emission from a Type II burst drops slowly in frequency as the shock moves away from the Sun into lower density regions at speeds of 400–2000 km s$^{-1}$. Type III bursts are generated by fast (2–20 keV) electrons from magnetic reconnection, typically due to solar flares. As the fast electrons escape at a significant fraction of the speed of light into the heliosphere along open magnetic field lines, they produce emission that drops rapidly in frequency. As discussed in §2.1, FARSIDE will also attempt to detect intense type II and type III emissions from other stars in the solar neighborhood.

***Acceleration at Shocks:*** Observations of CMEs near Earth suggest electron acceleration generally occurs where the shock normal is perpendicular to the magnetic field [Bale et al. 1999], similar to acceleration at planetary bow shocks and other astrophysical sites. This geometry may be unusual in the corona, where the magnetic field is largely radial. There, the shock at the front of a CME generally has a quasi-parallel geometry. Acceleration along the flanks of the CME, where the magnetic field-shock normal is quasi-perpendicular would seem to be a more likely location for the electron acceleration and Type II emission. The radio array needs ~2° resolution to localize these acceleration site(s), yielding the preferred geometry of the shock normal relative to the magnetic field direction for radio emission around CMEs.

***Electron and Ion Acceleration:*** Observations at 1–14 MHz made with the Wind spacecraft showed that complex Type III-L bursts are highly correlated with CMEs and intense (proton) solar energetic particle (SEP) events observed at 1 AU [Cane et al. 2002; Lara et al. 2003; MacDowall et al. 2003]. While the association between Type III-L bursts, proton SEP events, and CMEs is now secure, the electron acceleration mechanism remains poorly understood. Two competing sites for the acceleration have been suggested: at shocks in front of the CME or in reconnection regions behind the CME. For typical limb CMEs, the angular separation of the leading edge of the shock and the hypothesized reconnection region behind the CME is approximately 1.5° when the CME shock is 3–4 RS from the Sun.

***CME Interactions and Solar Energetic Particle (SEP) Intensity:*** Unusually intense radio emission can occur when successive CMEs leave the Sun within 24 hours, as if CME interaction produces enhanced particle acceleration [Gopalswamy et al. 2001, 2002]. Statistically associated with intense SEP events [Gopalswamy et al. 2004], this enhanced emission could result from more efficient acceleration due to changes in field topology, enhanced turbulence, or direct interaction of the CMEs. Lack of radio imaging makes it difficult to determine the nature of the interaction. Images with ~2° resolution would give Type II locations and permit identification of the causal mechanism and the relation to intense SEPs.

FARSIDE provides the requisite spatial resolution, as well as exquisite signal to noise and imaging capability (due to 8128 unique baselines) spanning the 1–14 MHz band observed by the Wind spacecraft and extending much wider at both lower and higher frequencies. It will be a uniquely powerful tool for understanding the complex interplay of CMEs, SEPs and associated shocks and particle acceleration.

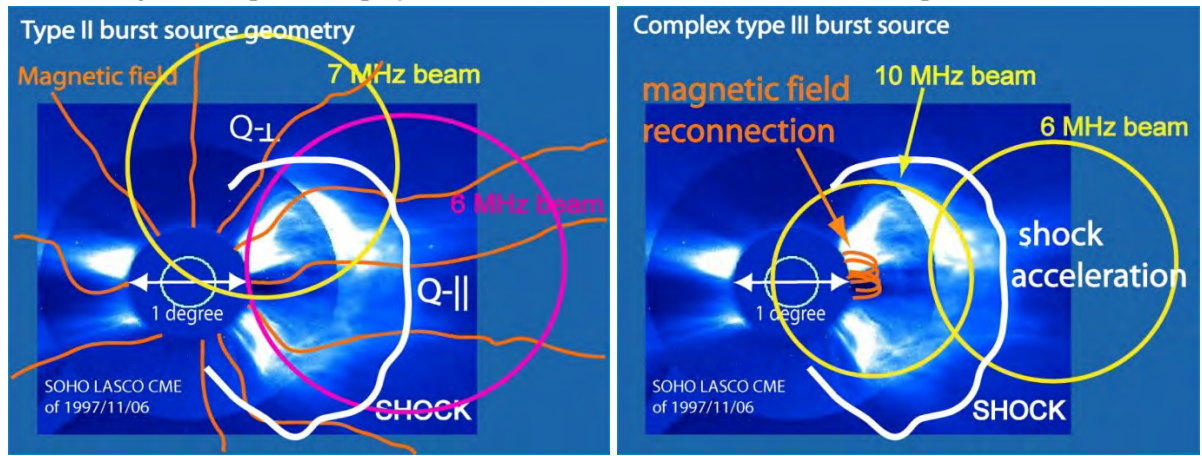

**Figure 2.3-1.** *Left:* Where on the shock does electron acceleration occur, yielding type II radio emission? *Right:* Are complex type III-L bursts produced by shock acceleration or reconnection?

### 2.3.2 *Lunar Quakes and Core with FARSIDE Optical Fibers*

Tectonic activity and internal structure of planets/moons are fundamental to understanding their formation, evolution and habitability. The moon is of particular interest, due to its close relation with the Earth [Canup 2012]. Four seismometers deployed during the Apollo missions provided fundamental constraints on the Moon activity and structure [Nakamura et al. 1982]. However, outstanding scientific questions, such as the mantle structure and seismic activity at the far side and the existence of a liquid core, remain unresolved. Answering these questions requires more seismometers, desirably dense seismic arrays, on the far side. Constrained by technical difficulties and cost, planetary seismology has mostly been focusing on single devices (e.g., Mars InSight) or a few stations (e.g., Apollo).

The ~50-km long fiber optic cables laid down by FARSIDE could enable the deployment of a 10-km aperture, 5000-sensor seismological observatory on the far side of the Moon, with minimum additional cost. The Distributed Acoustic Sensing (DAS) technology can turn every few meters of a 10's of km long optical fiber into a seismic sensor, by attaching an interrogation unit to one end of the fiber [Zhan 2019]. DAS works by shining a laser pulse into the fiber from one end and interrogating the "echo" of Rayleigh scattering from intrinsic fiber defects. If a fiber section is strained, the relative positions of the fiber defects within the section will change, and so will the corresponding backscattering, in both amplitude and phase. DAS measures the changes at a high sampling rate (e.g., 10 kHz) and converts them to strain measurements. The technology has proven to work well on Earth with either dedicated fiber optic cables or existing telecom fibers, without interfering data transfer over other strands of fiber in the same cable.

The seismological component of FARSIDE will require a few more strands of fiber in the cables connected in a continuous loop, and additional payload of DAS instrument at the FARSIDE Base Station. It can potentially share systems of power supply, environmental protection, and data processing/telemetry with the radio telescope component. Current commercial DAS instruments weigh about 50 kg and consume 200 W power, most of which is related to the computing units. A space qualified FPGA solution to the data processing could substantially reduce the weight and power consumption. It is thus a possible low-impact future augmentation of the FARSIDE baseline design.

The FARSIDE seismic network will answer two key science questions among many others:

***Deep moonquakes at the far side:*** The Apollo seismometers detected thousands of moon quakes at 700-1100 km depth from the lunar nearside, but much fewer from the far side [Nakamura 2005]. This lack of far-side events could be due to detection bias with all the sensors on the near side, or a true hemispheric difference in lunar deep structure and dynamics, which would have critical implications to lunar evolution [Qin et al. 2012; Laneuville et al. 2013] (Figure 2.3-2A). It is well known that shallow lunar structure has clear hemispheric asymmetry, but the depth extent of which is still unclear. To maximize the detection capability, we can apply array-processing techniques to the FARSIDE seismic network and enhance data signal-to-noise ratios by tens of times.

***Size and properties of the lunar core:*** Reflections from the lunar core-mantle boundary have been difficult to observe, due to scattering in the highly heterogeneous lunar crust [Weber 2011]. Therefore, there are still large uncertainties on the lunar core size and properties. Does the Moon have liquid outer core and a solid inner core like the Earth? Constraining the structure of core is almost always the top goal of planetary geophysical program due to its importance to understanding planetary evolution and habitability. Previous studies on Earth and synthetic simulations for the Moon have demonstrated that the 10-km aperture FARSIDE array would greatly enhance the lunar core phases hidden in scattered waves and provide a clear picture of the lunar core (Figure 2.3-2B).

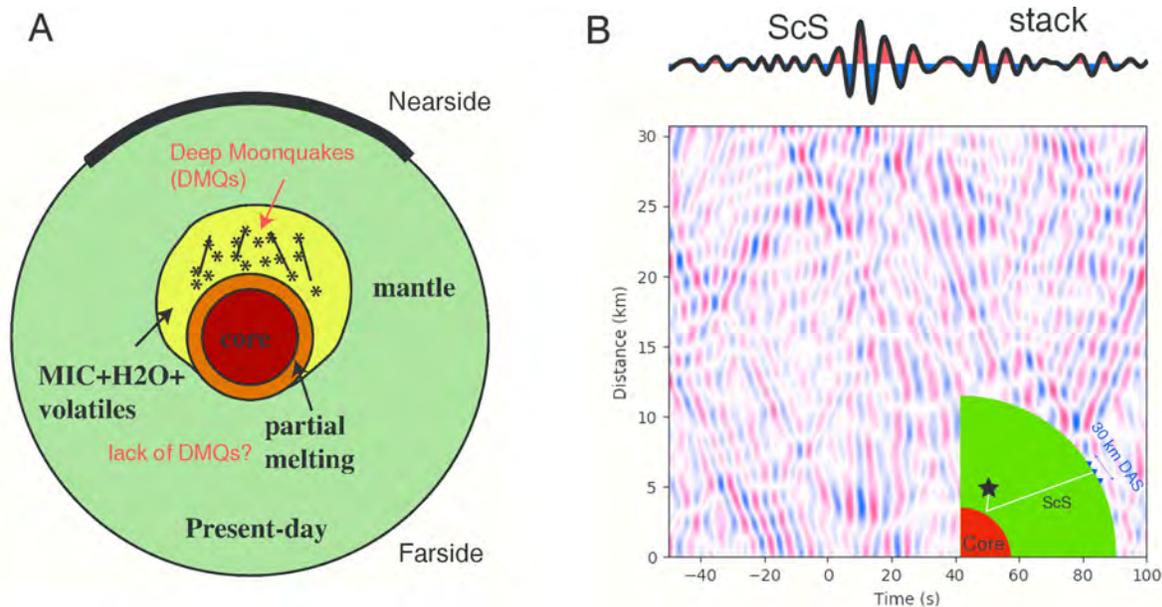

**Figure 2.3-2.** Hypothetic hemispherical lunar structure and a synthetic example of DAS array technique. (A): A schematic representation of the hemispherical lunar structure. Adapted from Qin et al. [2012]. (B): A synthetic example of retrieving core phase ScS from a 30-km long DAS array. The strongly scattered seismic waves recorded over the DAS array completely mask the weaker ScS core phase (ray path shown in the inset), but array stacking enhances the signal as shown in the top panel. Time zero is the ScS arrival time predicted by the 1D model [Garcia et al. 2011].

### 2.3.3    *Outer Solar System Planets*

In 1955, the planet Jupiter was serendipitously detected to be a source of bright, highly variable radio emission at decametric wavelengths [Burke and Franklin 1955]. This radio emission was manifested as intense, highly polarized bursts, so bright as to often outshine the Sun at these low frequencies (< 40 MHz). Bursts were only detectable over specific ranges of rotational phase of the planet, with certain flavors also found to be strongly modulated by the orbit of the volcanic moon Io. This discovery revolutionized our understanding of planetary magnetospheric physics, providing the first direct confirmation of the presence, strength and extent of the Jovian magnetosphere. In particular, the radio bursts, generally attributed to electron cyclotron maser emission, were found to be produced at the electron cyclotron frequency, $v_{ce} \approx 2.8 \times 10^6 \; B$ Hz, where is the magnetic field strength at the source of the burst, and thus enabled direct determination of magnetic field strength.

All the magnetized planets in our solar system, including Earth, have since been found to produce similar bright coherent radio emission at low frequencies, predominantly originating in high magnetic latitudes and powered by auroral processes. It is notable that only Jupiter, with a maximum polar magnetic field strength of 14 Gauss, produces radio emission that can penetrate through the Earth's ionosphere. All the other planets (and Ganymede), have magnetic fields of <2 Gauss, requiring a space-based observatory for detection. For example, the radio emissions of Saturn, Neptune and Uranus were initially detected in-situ by Voyager during fly-by [Zarka 1998].

These auroral processes are driven by a) magnetic reconnection between the planetary magnetic field and the magnetic field carried by the solar wind (e.g., Earth, Saturn, Neptune and Uranus), b) the departure from co-rotation with a plasma sheet residing in the planetary magnetosphere (e.g., Jovian main auroral oval) or c) interaction between the planetary magnetic field and orbiting moons (e.g., Jupiter-Io current system) [Zarka 1998 and references therein]. As well as providing diagnostic information on the presence, strength and extent of planetary magnetospheres, the detected radio bursts are the only means to accurately determine the rotation period of the planetary interior for the gas giants.

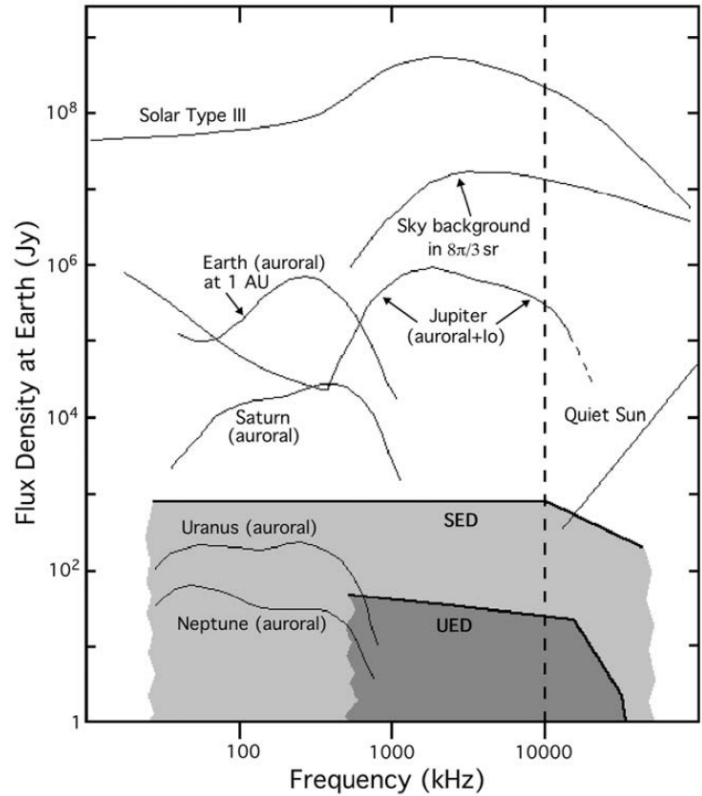

**Figure 2.3-3.** The average radio power as a function of frequency of solar system low frequency radio emission, as received at Earth (with the exception of Earth, which is normalized to a distance of 1 AU). This includes auroral emission from all the magnetized planets, and lightning from Saturn (SED) and Uranus (UED). The dashed vertical line is the ionospheric cutoff, i.e., the frequency below which a space-based radio telescope is required for detection. Note that FARSIDE sensitivity is below the lower edge of the plot for all frequencies, within 60 seconds of integration time. Figure from Zarka [2004].

FARSIDE will be the first space-based radio telescope with sufficient sensitivity for detection of radio emission from all the magnetized planets from Earth-orbit. Monitoring the impact of the variable solar wind on the auroral radio flux density at the planets, including during interplanetary shocks from CMEs, is one application. Long-term monitoring of the Kronian radio emission period, modulated by the solar wind, will refine measurement of the true planetary rotation period [Zarka et al. 2007]. Neptune and Uranus have not been observed at radio wavelengths since Voyager.

In addition, Saturn and Uranus have been detected by Voyager to produce radio emission associated with atmospheric lightning [Zarka et al. 2004] (Figure 2.3-3). This radio emission can potentially inform on atmospheric dynamics and composition, when compared to optical imaging of cloud data. These emissions are also easily detectable by FARSIDE.

Finally, we note that FARSIDE would offer the unique possibility of searching for radio emission from large bodies beyond Neptune out to 100s of AU. This includes, but is not limited to, the putative Planet 9 .

The expected flux density of Planet 9 is ~10 mJy below 1 MHz, if a significant planetary magnetosphere is present.

### 2.3.4    *Sounding of the Lunar Sub-surface*

FARSIDE has the potential to sound the mega-regolith and its transition to bedrock expected at ~2 km below the surface [Tauscher et al. 2018]. The Lunar Radar Sounder (LRS) onboard the KAGUYA (SELENE) spacecraft has provided sounding observations of the lunar highlands [Chianese et al. 2019] and found potential scatterers in hundreds of meters below the subsurface. However, the results are inconclusive due to surface roughness. FARSIDE, by virtue of being on the surface, would not be affected by roughness. Data from a calibration beacon in orbit could be synthesized to identify deep scatterers and the transition to bedrock at km depths by virtue of the low frequencies, which are significantly more penetrating. Deep subsurface sounding can also be performed passively using Jovian bursts from 150 kHz to 20 MHz [Bradenberger et al. 2019]. The array covers a 10 km × 10 km area on the lunar highlands which could provide a three dimensional image of highland subsurface structure.

### 2.3.5    *Tomography of the Local ISM*

Free-free absorption can be used to easily identify the presence of HII regions, backlit against the much brighter Galactic synchrotron emission, using ground-based low frequency radio telescopes at a few 10s of MHz. However, by ~1 MHz the entire sky becomes optically thick to free-free absorption by the warm ionized medium (WIM). The full three-dimensional structure of the warm interstellar medium (WIM) in the solar neighborhood can be inferred through mapping the emissivity as a function of frequency over a wide range of frequencies with FARSIDE [Reynolds 1990]. Simultaneously, the emissivity of the synchrotron emission in the Galactic disk and halo can be measured, providing key insight to the low energy tail of the cosmic-ray energy distribution.

Finally, the use of rotation measure synthesis [Brentjens and de Bruyn 2005] can be used to map the very near interstellar plasma to the solar system, potentially probing the interstellar medium within 1 pc.

# 3    INSTRUMENT

## 3.1    Overview

As a low frequency radio observatory (100 kHz–40 MHz) targeting faint signals, the FARSIDE array is required to be above the Earth's plasmasphere and to be shielded from radio frequency interference (RFI). Ground-based radio astronomy is limited to > 10 MHz, due to attenuation in the ionosphere. Moving above the atmosphere opens an additional 2 orders of magnitude in frequency range below which observations become limited by the plasma density of the solar wind at Earth orbit. However, the sensitivity of an Earth-orbiting radio telescope is increasingly limited below ~1 MHz. This is partially due to the presence of anthropogenic radio frequency interference (RFI) and the intense auroral kilometric radiation (AKR) produced by Earth. More fundamentally, the system noise below 1 MHz in Earth orbit is limited by the electrons in the solar wind colliding with the antenna and inducing currents [Meyer-Vernet and Perche 1989]. Together, these factors prevent sensitive searches for exoplanet radio emission below 1 MHz. By contrast, the lunar farside provides a unique environment for radio astronomy within the inner solar system, with > 100 dB attenuation of both RFI and AKR. Most importantly, a plasma cavity exists on the surface, particularly on the night time side, that provides sufficient isolation for sky noise dominated observations down to ~200 kHz.

Provided this environment, unique in the inner solar system, the design of the radio astronomical array is driven by sensitivity and spatial resolution, both specified in Table 3.1-1. The sensitivity is determined by the number and type of antenna used while the spatial resolution is determined by the distribution of baselines (i.e., pairwise antenna separation vectors). The choice of antenna for FARSIDE is an electrically short dipole, which is simple and commonly applied to low frequency observations. Given the antenna effective area and noise temperature (§3.2), 128 antennas in each band (100 kHz–2 MHz and 1–40 MHz) will be needed to achieve the sensitivities in Table 3.1-1. The spatial resolution in Table 3.1-1, can be achieved with maximum baseline separations of 10 km.

**Table 3.1-1.** Baseline FARSIDE specifications.

| Quantity | Value |
|---|---|
| Antennas | 128 × 100 m length dipoles (100 kHz – 2 MHz), 128 × 5 m length dipoles (1-40 MHz) |
| Frequency Coverage | 100 kHz – 40 MHz (1400 × 28.5 kHz channels) |
| Field of View (FWHM) | > 10,000 deg$^2$ |
| Spatial Resolution | 10 degrees @ 200 kHz / 10 arcminutes @ 15 MHz |
| Antenna efficiency | $6.8 \times 10^{-6}$ @ 200 kHz / $9.5 \times 10^{-5}$ @ 15 MHz |
| System Temperature[a,b] | $1.0 \times 10^{6}$ K @ 200 kHz / $2.7 \times 10^{4}$ K @ 15 MHz |
| Effective Collecting Area[c] | ~ 12.6 km$^2$ @ 200 kHz / 2,240 m$^2$ @ 15 MHz |
| System Equivalent Flux Density (SEFD) | 230 Jy @ 200 kHz / $2.8 \times 10^{4}$ Jy @ 15 MHz |
| 1σ Sensitivity[b] (60 seconds; bandwidth = **ν**/2) | 93 mJy @ 200 kHz[d] / 1.3 Jy (1.2 K) @ 15 MHz |
| 1σ Sensitivity[b] (1 hour; bandwidth = **ν**/2) | 12 mJy @ 200 kHz[d] / 170 mJy (160 mK) @ 15 MHz |
| 1σ Sensitivity[b] (1000 hours; bandwidth = **ν**/2) | 230 μJy[e] @ 200 kHz[d] / 3.8 mJy (5.2 mK) @ 15 MHz |

[a] *System temperature includes contribution from the sky and ground due to the absence of a ground screen.*
[b] *These values have been updated from the Astro 2020 report due increased fidelity in the front-end design (see §3.5).*
[c] *Effective area is impacted by loss of gain into the ground due to absence of a ground screen. Antenna efficiency not included.*
[d] *Sensitivity calculations at 200 kHz assume night time conditions.*
[e] *Deep confusion-free integrations are possible < 3 MHz due to the absence of extragalactic sources.*

FARSIDE would consist of a central Base Station and 128 antenna nodes distributed across a 10 km diameter area (Figure 3.1-1). The nodes are distributed along eight independent spokes (16 nodes per spoke), with a science tether connecting the nodes in each spoke, distributing power and providing a signal path back to the Base Station. The eight spokes are deployed to form four petals with a deliberate asymmetry to improve imaging performance, and to minimize rover travel on the lunar surface during deployment. For each petal, the deployment rover (Figure 3.1-2) would carry a set of antenna nodes out from the base station, unreeling the tether, before returning to base along a different path, continuing to unreel the tether. The

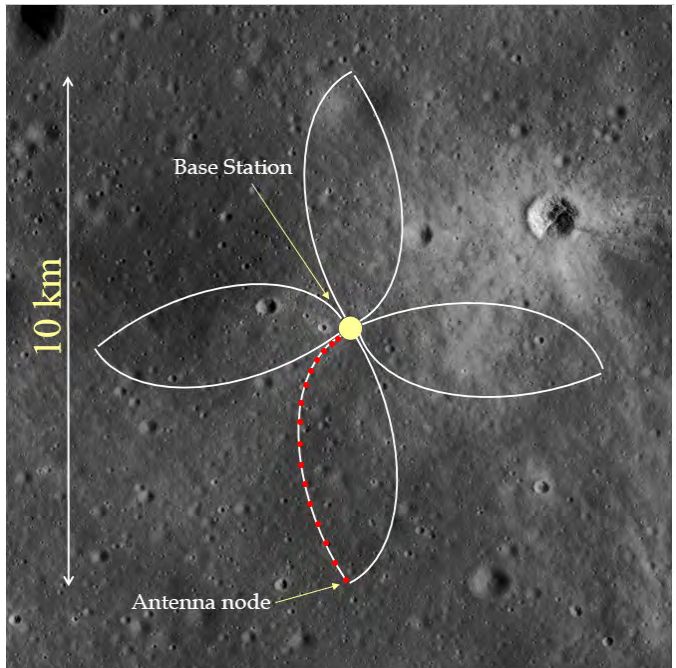

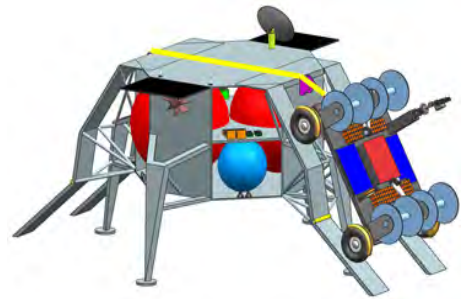

Base Station would provide power, signal processing, and telecommunications back to Earth (via a relay). Each node consists of two dipoles, one built into the tether for low frequencies and a short dipole, orthogonal to the tether, for the higher frequencies. Each antenna has a preamplifier, sending the two signals back to the Base Station via optical fibers as an analog signal.

The antenna nodes are powered in series similar to the scheme used for transoceanic cables and airport runway lights, so only two power conductors are needed (although six conductors are used for redundancy). All of the antenna nodes draw essentially the same power, since they are in the same environment and are identical circuits. The base station provides a constant current of about 20 mA, and needs to have sufficient voltage compliance to accommodate the changing voltage drop in the tether (due to resistance changes from temperature) and the changing load of all the nodes (also primarily due to temperature changes). Each node has a small transformer which provides galvanic isolation between the tether power and the internal circuitry, and provides continuity in the event of a node failure. The voltage drop across the transformer primary varies in direct proportion to the load current on the secondary. A similar system is used for LED airport lighting along taxiways and runways, where the overall loop length is kilometers.

## 3.2    Short Dipole Antennas

The receiver does not use resonant antennas, rather it uses electrically short dipole antennas following the proven design approach of Ellingson [2005]. The electrically short antenna only needs to be efficient enough to be sky-noise dominated. This is achieved by having a highly resistive input impedance in the front-end amplifier that can be chosen to achieve sky-noise limited performance over a wide band. The antenna impedance is composed of the radiation resistance and the loss resistance. The radiation resistance due to the length of the antenna and surrounding dielectric, the loss resistance due to dielectric losses and finite conductivity of the elements, and the reactance resulting from the antenna length and wire diameter were calculated using the NEC4.2 numerical model [Burke and Poggio 1992] assuming the lunar regolith's relative dielectric constant of $\varepsilon_r = 3 + 0.005j$ from [Carrier et al 1991]. For both frequency bands, the front-end impedance is loaded with a resistor (1 MΩ) to achieve sky noise limited performance over a wide band. The effective collecting area and system temperatures estimated from these calculations (Figure 3.2-1) are the basis for the sensitivity estimates (§3.5) presented in Table 3.1-1.

Since the antenna is so close to the regolith, in terms of wavelength, the use of a ground screen (as used at higher frequencies in observatories such as OVRO-LWA and LOFAR) is impractical. As a consequence, the majority of the antenna beam is pointed into the regolith. The electrical performance of a wire antenna near, on, or below, a dielectric surface has been studied in detail by Rutledge and Muha [1982] and is fully accounted for in the NEC4.2 simulations. The main differences with a free-space dipole is that it radiates more power into the dielectric than the air (or space) by $\varepsilon_r^{3/2}$, a power ratio of about 5 (7 dB) for typical regolith $\varepsilon_r = 3$. The effective height of the antenna is 73% of the free space effective height in the zenith direction. FARSIDE uses 2.5-meter dipole arms (5-meter length total) for the high frequency band (1–40 MHz), producing an effective height of approximately 1.8 meters. The low frequency antenna (100 kHz – 2 MHz) is embedded in the tether and has 50 meter long arms (100-meter length total). As with the high frequency band, the regolith loading reduces the effective height to about 36 meters. The band of operation for each antenna is limited by sensitivity on the low frequency end and by multi-lobe radiation pattern of the antenna in the high frequency end, which becomes impractical to calibrate. While dual polarization operation of the array is not part of the current baseline design, it is anticipated that full bandwidth (100 kHz–40 MHz) sky noise-limited sensitivity will be possible for each polarization with completion of ongoing design/development work. This will enable circular polarization to be used as a confirmation of exoplanet radio emission and reduce the estimated integration times for all science cases by ~2.

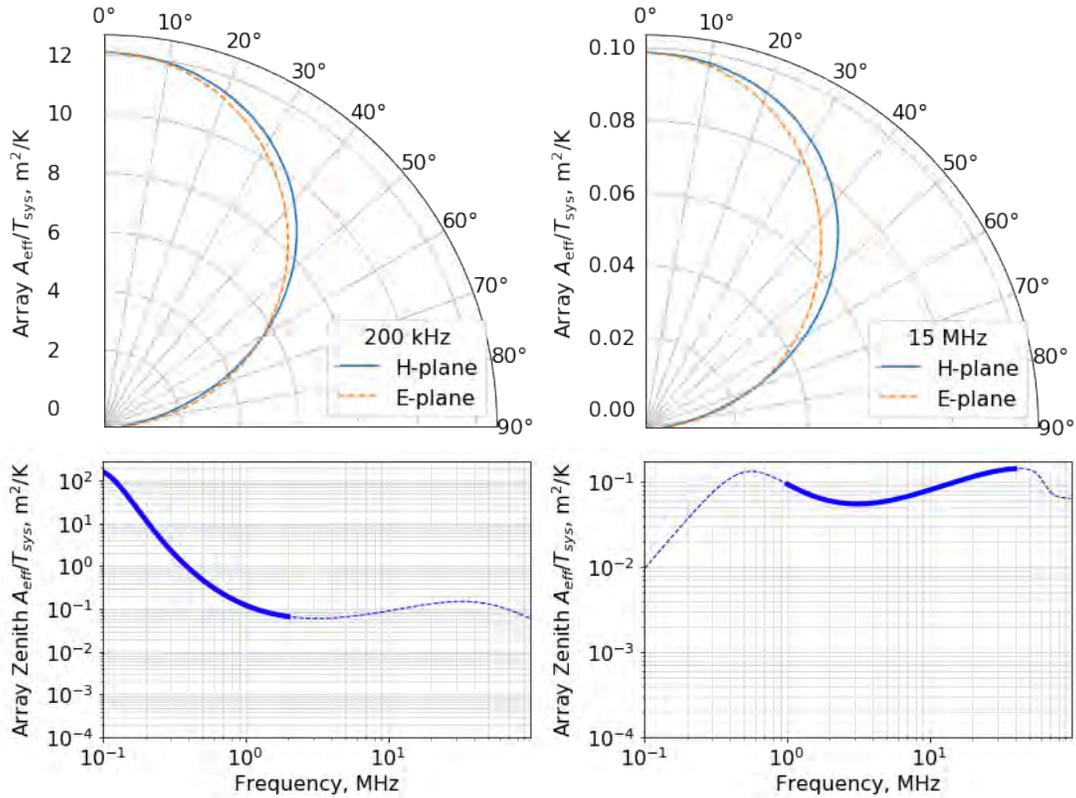

**Figure 3.2-1.** *Top*: Effective area/system temperature as a function of zenith angle at 200 kHz (*left*) and 15 MHz (*right*) for the 128-antenna low and high frequency array, respectively. The antenna beam patterns are obtained from NEC4.2 simulations, which fully account for the effects of the dipole lying directly on the regolith. The system temperature estimates are presented in §3.5. *Bottom*: the effective area/system temperature as a function of frequency are shown for the low frequency band (*left*) and high frequency band (*right*) arrays. The band of operation is shown as a thick line for each case.

Each Antenna Node (left panel of Figure 3.2-2) consists of a small ($5 \times 5 \times 10$ cm) aluminum housing containing two receiver channels, each comprised of a high input impedance low noise preamplifier using an operational amplifier driving a laser diode feeding an optical fiber back to the Base Station. The low-frequency dipole is embedded in the tether and the high-frequency dipole is laid out at right angles to the tether by two STACER deployable elements. To save tether mass, all the nodes in a "petal" are connected in series along the tether so there are only two power wires in the tether. All elements for each band are oriented in the same direction by bending the parts of the tether with the low-frequency antenna element relative to the petal pattern in the right panel of Figure 3.2-2.

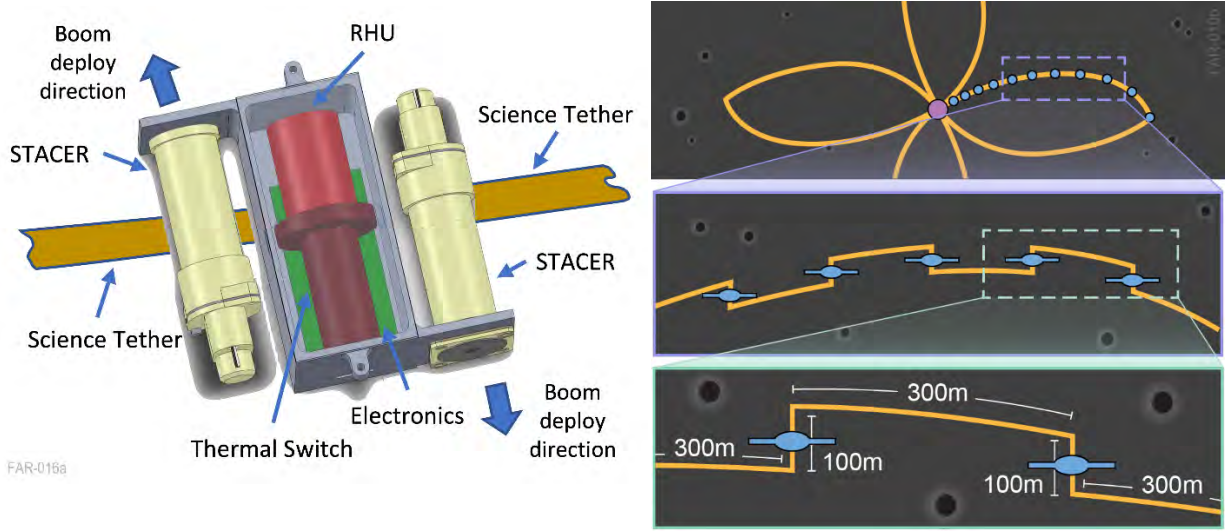

**Figure 3.2-2** *Left*: Mechanical design of the FARSIDE antenna node. The low frequency dipole is in the tether while the high frequency STACER dipole deploys orthogonally. *Right*: Zoom in of the antenna layout. The low frequency antennas, embedded in the tether, are all aligned vertically across the array. This is achieved on each petal using the zigzag design for the portion of the tether containing the antenna. The high frequency STACER dipole is deployed orthogonal to the low frequency dipole in the tether.

A ~300 V voltage power source at the base station injects a constant current and a switching converter at each node produces the voltages needed by the antenna front-end electronics. The AC input impedance of the power supply is nearly infinite at the RF frequencies of interest, so it does not short the antenna. Periodically inserted series inductors in the power tether wires prevent interactions with the antenna. The node is powered by a conventional AC/DC rectifier followed by a LT3042 low drop out linear regulator that provides over 70 dB of isolation below 3 MHz and above 3MHz, where the isolation drops to only 55dB, conventional LC filters provide additional isolation. No switching regulators are used in the node. The base station AC current source is filtered to eliminate any in-band harmonics or transients. The final AC frequency is the subject of future trade studies with respect to transformer losses and efficiency, but is expected to be in the range 100 Hz – 10 kHz. The tether will be broken up with chokes presenting a high RF impedance to prevent interaction with the antenna. The design work for this is ongoing.

## 3.3    Receiver

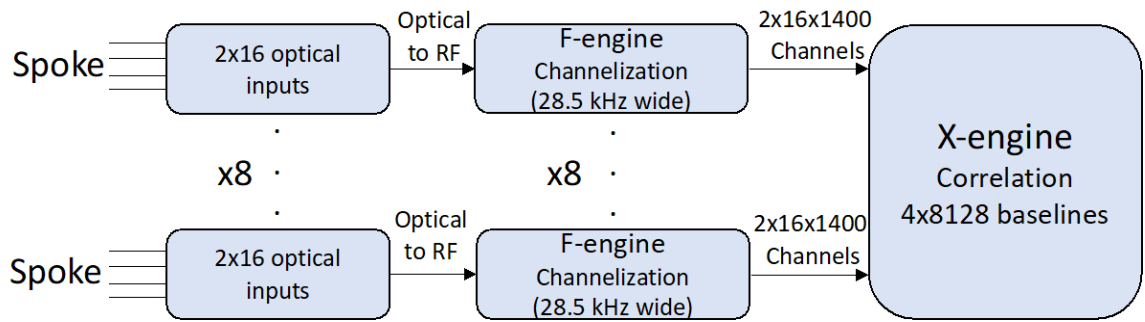

**Figure 3.3-1**. F-X correlator diagram. Each spoke has 16 nodes, with 2 antennas each. The signals arrive via optical fiber and are converted back to radio frequency (RF) and channelized via the FPGA-based F-engine. A subset of the channels is fed for correlating all baseline pairs via the X-engine.

The Base Station houses the FX correlator (Figure 3.3-1) - each of the eight spokes feeds an F-engine board which receives the 32 optical signals (two fibers per antenna node) and performs the frequency channelization in space qualified Field Programmable Gate Arrays (FPGAs). The outputs of those boards are corner-turned to X-engine boards that perform full cross-correlation of all antennas, also in FPGAs. The integrated output of the correlators is passed to the JPL-developed, radiation-tolerant, FPGA-based, Miniaturized Sphinx onboard computer and stored for relay to Earth. 192 GB of storage is provided, about three days at full rate. The Sphinx is also responsible for system management, engineering housekeeping telemetry, and control functions.

## 3.4    Signal Processing and F-X Correlator

FARSIDE would collect full cross-correlation data every 60 seconds, in 1400 channels of 28.5 kHz width each, for a total data rate of 65 GB per 24-hour period. All visibility data output by the X-engine would be transferred via the Lunar Gateway (or orbiting relay) to Earth. A snapshot image of the ~10,000 deg$^2$ area within the half-power primary beam of the antennas would be produced for each of the 1400 channels in each band, as well as a combined full-band image. Dynamic spectra at the location of every stellar/planetary system within 25 pc would be produced from these data to search for Type II/III radio bursts and planetary auroral emission. In addition, data from each lunar day would be combined via the technique of m-mode analysis [Shaw et al. 2014] to produce deep all-sky images that would precisely map the synchrotron and absorbed free-free emission in our Galaxy, as well as providing the best means for deep searches for the median flux densities of exoplanets. The use of snapshot all-sky imaging together with deep integration all sky imaging via m-mode analysis has already been demonstrated for the OVRO-LWA, albeit at higher frequencies than accessible to FARSIDE [Anderson et al. 2018; Eastwood et al. 2018]. Simultaneous auto-correlation power spectra for each antenna would be collected every 60 seconds, which together with the m-mode foreground maps, would deliver the radio spectrum of the Dark Ages 21-cm signature.

## 3.5    Sensitivity

The array sensitivity ($\Delta S$), or minimum detectable flux, provided in Table 3.1-1, is given in terms of the array's system equivalent flux density (SEFD) via the relation $\Delta S = \text{SEFD}/\sqrt{\Delta f \Delta t}$, where $\Delta t$ is the integration time and $\Delta f$ is the bandwidth. The SEFD for the interferometer array depends on the array effective area $A_{eff}$ and system temperature $T_{sys}$ according to:

$$SEFD = 2k_B \frac{T_{sys}}{A_{eff}}, \qquad\qquad (\text{Eq. 3.5-1})$$

where $k_B$ is Boltzmann's constant. Note that the effective area for the interferometer array $A_{eff}$ is given in terms of the effective area for a single dipole $A_{eff,dipole}$ and the number of antennas $N$ via $A_{eff} = A_{eff,dipole}\sqrt{N(N-1)}$. The exoplanet observation objectives drive the optimization of the effective area over temperature (Figure 3.2-1), which requires a balance between antenna size, input impedance, and amplifier noise characteristics. The Dark Ages 21-cm signature objective drives the optimization of sky-noise limited observations.

The antenna effective area is given by the dipole length and antenna impedance. This is determined from NEC4.2 simulations, which take into account the effects of the wires resting directly on the regolith (§3.2). The front-end amplifier is a discrete FET feeding low noise operational amplifier (OpAmp) and its input impedance, as seen from the antenna terminals, is resistive and settable (1 MΩ for both bands). We assume the OpAmp has voltage noise of $v_n$=2 nV/Hz$^{1/2}$ and current noise $i_n$=1.3 fA/Hz$^{1/2}$, which is consistent with previously built low noise OpAmps [Jefferts and Walls 1989]. The full noise model of the OpAmp was used, including contributions from the current noise and operating temperature noise, and the total noise temperature of the amplifier is dominated by the voltage noise term in the frequency bands used for this study. The 1/f noise of these operational amplifiers is subdominant above 10 kHz so flicker noise is not of concern for this application.

The total system temperature has two contributions: the antenna temperature $T_{ant}$, which is given by the sky and regolith temperatures, and the front-end amplifier $T_{amp}$. Although technically each element of the signal chain contributes to the receiver noise, the front-end amplifier dominates by far provided the system is designed with sufficient gain in the first stage, which is standard practice.

The antenna temperature assumes the average sky temperature measurements at low frequencies [Novaco and Brown 1978], a lunar night-time regolith temperature of 100 K [Heiken et al. 1991]. The total antenna temperature results from integrating the sky and regolith temperatures with the dipole-on-soil beam pattern [Rutledge and Muha 1982] and multiplying it by the antenna efficiency. Because of this, the total antenna temperature at 15 MHz results in 23,000 Kelvin rather than the sky noise of 100,000 Kelvin. At 200 kHz, the noise equivalent temperature would be $1.3 \times 10^6$ Kelvin compared to the expected sky noise of $7 \times 10^6$ Kelvin. Note that the values in of effective area and system temperature shown in Table 3.1-1 do not include the antenna efficiency, but this term is included in the estimates of $A_{eff}/T_{sys}$, the SEFD, and flux sensitivity. The antenna efficiency is dominated by the load efficiency $\epsilon = 4R_A R_L/|Z_A + Z_L|$ were $R_A$ is the antenna resistance, $Z_A$ is the complex antenna impedance, $R_L$ is the antenna resistance, and $Z_L$ is the complex antenna impedance. The antenna impedance values are obtained from NEC4.2 simulations while the load impedance is set to be predominantly resistive with $R_L = 1$ MΩ. While the resulting antenna efficiency is low (Table 3.1-1), the observations are sky noise dominated. The sky noise dominance, defined as $T_{ant}/T_{noise}$, where $T_{noise}$ are all contributions to the system temperature except for the sky, are shown in Figure 3.5-1. With the design presented here, FARSIDE would provide sky noise-limited observations.

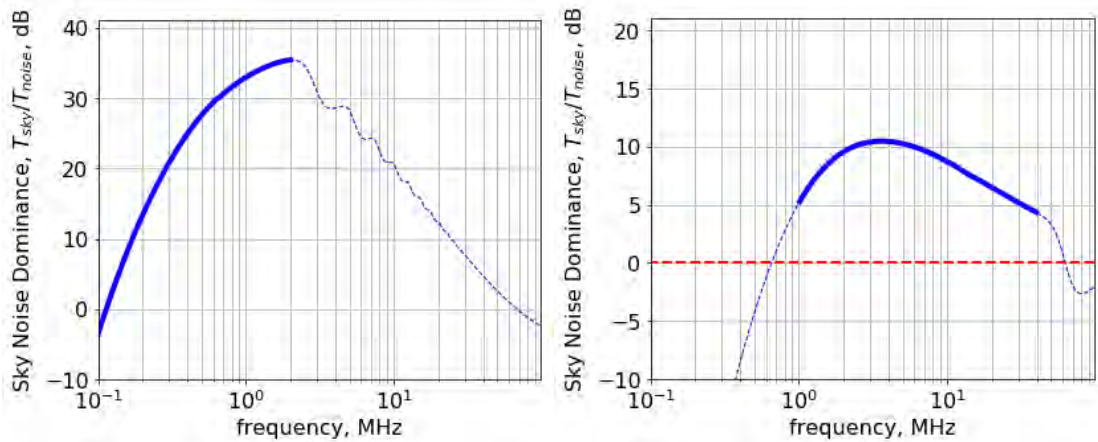

**Figure 3.5-1.** Sky noise dominance as a function of frequency are shown for the low band (*left*) and high band (*right*) antennas. The red dashed line at 0 dB indicates where sky noise would be equal to other sources of noise. The band of operation is shown as a thick line for each case.

The bandwidth $\Delta f$ is 28.5 kHz and the integration time for each acquisition is 60 seconds. For a single snapshot, this provides a gain $\sqrt{\Delta f \Delta t}$ =31 dB per snapshot and stacking multiple snapshots and frequency bands further increases the sensitivity as shown in Table 3.1-1.

### 3.6    Calibration

The calibration of the array will be carried out using astronomical sources, as is standard practice. However, to meet the Dark Ages 21 cm signature objective, there is a requirement to obtain a relative antenna beam pattern knowledge to a level of < 50 dB (§1.1, STM). Knowledge of the antenna beam will be crucial for separating the beam-averaged foreground from the 21-cm signal. To achieve that level of relative beam pattern calibration, an external source will be needed. Possible implementations being evaluated include a dedicated transmitter mounted on a SmallSat. This would be the preferred implementation since the beam pattern calibration requires that the array is illuminated from all directions relevant to observations which will require a precessing orbit. The orbiting beacon requirements will be the subject of further investigation.

# 4  DESIGN REFERENCE MISSION

## 4.1  Overview

The mission architecture includes three components: a commercial lander (Figure 4.1-1), a FARSIDE base station which is part of the commercial lander, and a rover which is used to deploy the 128 receiving antenna nodes. The commercial lander is separate from the base station, except for providing structure and the ride to the Lunar surface. The base station uses nuclear power to allow operation through the two week long Lunar night, with some solar panels for additional power. The rover is solar powered, and only operates during the Lunar day, with radioisotope heater units (RHU) to stay warm during the night.

The commercial lander carries the entire package to the Lunar surface, providing power and communication until landing, at which time the base station starts up. The rover rides on top of the commercial lander, and drives down to the surface via ramps and an assist winch. Once the rover is deployed on the surface, and the base station is in operation, the commercial lander can shut down or continue activities, assuming that it meets electromagnetic interference requirements.

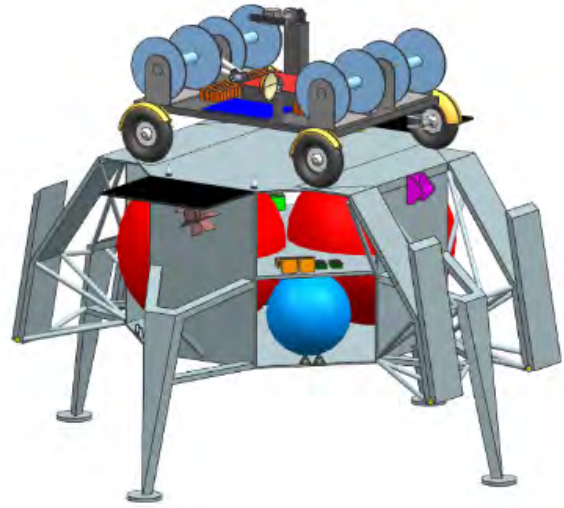

**Figure 4.1-1.** The commercial lander includes the base station hardware, and carries the tether deployment rover on top. The overall assembly is ~4 meters in diameter, stowed, and 6 meters tall.

After landing, the rover is deployed, and lays eight tethers with small receiver nodes over a 10 km extent on the Lunar surface. This is described in more detail in §4.3

### 4.1.1  Baseline concept of operations

The Base Station begins science operations seven months from Launch, this allows one month for the commercial lander cruise and landing on the Lunar surface, and 6 months for the deployment of the antenna array.

There is a desire to conduct science observations almost continuously, including through the lunar night, generating a total of 7270 GB of science data per lunar night (520 GB per 24-hour period). This would require simultaneous science ops and communications which requires a lot of power, more than can be produced by the two eMMRTGs in the base station. One strategy considered was to store the large (Terabyte) amount of data during the night and transmit during the day (when Solar power is available). This data volume is not currently available as a space qualified component/assembly, and the cost and risk of developing it was assessed as too great. An alternate strategy, also discarded, was to store less information, 3 days (fitting within currently available space-qualified hardware), and do simultaneous communications and observation, this approach exceeds the power generation capability of the eMMRTGs. The selected option is to alternate between science and downlink. With baselined telecom design and assumptions about Gateway's link capabilities, the downlink rate is roughly equivalent to the data collection rate, i.e. 24 hours of science +30% overhead takes 24 hours to downlink. This 50% duty cycle is feasible with two eMMRTGs by draining a battery during science and recharging during downlink. The overall system is flexible, so a tradeoff between observation time and data transmission time can be made during the mission by changing the number of channels being processed and transmitted.

## 4.2  Proposed Architecture and Array Configuration

FARSIDE would consist of a central Base Station and 128 antenna nodes distributed across a 10 km diameter area (Figure 4.2-1). The nodes are distributed along 8 independent tethers, which are deployed to form 4 petals with a deliberate asymmetry to improve imaging performance. For each petal, the deployment rover would carry a set of antenna nodes out from the base station, unreeling the tether, before returning to base along a different path, continuing to unreel the tether and attached nodes. The Base Station would provide power, signal processing, and telecommunications back to Earth (via a relay). The individual antenna nodes are a crossed dipole with preamplifier, sending the two signals back to the Base Station via optical fibers as an analog signal. The antenna nodes are powered in series similar to the scheme used for transoceanic cables, so only two power conductors are needed (although six conductors are used for redundancy).

### 4.3 The Lunar Environment

For the purposes of this study, we have assumed that the array will be positioned somewhere on the farside of the Moon, in an area of mid latitudes with geology that is stable.

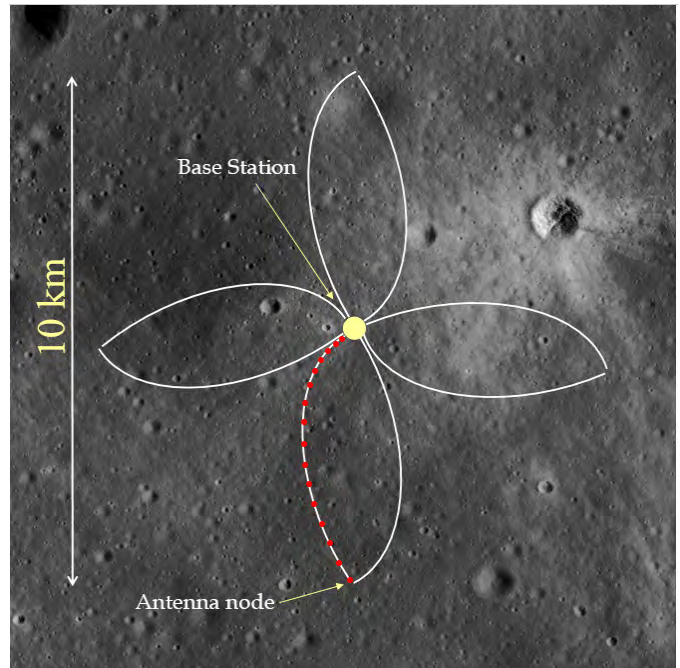

**Figure 4.2-1.** FARSIDE will consist of 128 antenna nodes deployed by a rover from a central base station, arranged in a petal configuration.

The receiving array should be no closer to the poles than 60° latitude, and no farther from the anti-Earth longitude than 60 degrees. This minimizes the effect of RF interference from the Earth propagating as a surface wave around the limb [Bassett, et al., submitted 2019]. Another reason to be in mid latitudes is that we wish to avoid areas where there is significant ice in the regolith or subsurface. Ice has significantly different electromagnetic properties than the relatively stable lunar regolith. While the regolith will change temperature (changing the EM properties) it is repeatable. In polar regions the ice/water content may change in an unpredictable way. Choosing an array location away from the poles reduces this contribution to system calibration uncertainty.

We further assume that daytime surface temperatures will be high (> 70C) and night time temperatures will be low (100K, −173C). The initial concept design uses a combination of Radioisotope Heating Units (RHU) to provide heat during the night, radiators to reject heat during the day, and passive thermal switches to disconnect the radiator during the night.

We have assumed a total dose of 30 kRad for the entire mission.

### 4.4 Lunar Gateway and Commercial Lunar Landers: Assumptions, Interfaces, Services

The study assumed that the mission would be designated as a Class B[1] mission for the risk posture (high national significance, design life 2–5 years, and high to medium cost—comparable to the Mars rover missions). This leads to a basic system architecture of dual string redundancy with cold standby.

We further assumed that the launch date would be January 2028, the study did not identify any specific launch date sensitivities: Launch opportunities to the Moon are frequent, unlike planetary exploration where launch periods may be only a few weeks every few years.

We have assumed that a commercial provider will deliver the lander and rover to the far side lunar surface, with all cruise operations and navigations support included in commercial provider costs, which are not included in the mission budget. During cruise, all communications to the lander and rover will be handled by the commercial carrier with a simple power and data interface. During cruise, the FARSIDE Ground system will receive spacecraft health and safety data from carrier provider and we will send commands to and through the carrier provider via some standard method such as IP Sockets.

---

[1] NPR 8705.4 Risk Classification for NASA Payloads

On the surface we will operate the rover and base station (mounted on the lander), which can operate entirely independent of the lander and each other, having their own power and RF communications capability. The commercial lander may be turned off after deployments are completed or the commercial provider may continue monitoring. We will impose EMI compatibility requirements on the commercial lander if it continues operation.

**Table 4.4-1.** Communications back to Earth will be conducted via the Lunar Gateway.

| Assumed Lunar Gateway Communications Support ||
|---|---|
| Rover forward link (Ku-band) | 16 kb/s |
| Rover return link (Ka-band) | 5 MB/s |
| Lander forward link (Ku-band) | 16 kb/s |
| Lander return link (Ka-band) | 10 Mb/s |

We assume that the landing site and rover path way is flat and free of obstacles larger than one meter. The one meter dimension is driven by current LRO imagery resolution. It is assumed that the rover will be able identify and drive around smaller obstacles, and that an area at least 10km in diameter (plus the landing position uncertainty) can be identified as a landing site.

We have assumed that communications back to Earth will be conducted via the Lunar Gateway, as described in the Gateway Architecture document. The Gateway is presumed to be in a Near Rectilinear Halo Orbit (NRHO) favoring the south Lunar pole as described in [Zimovan, et al. 2017] The communications system on Lunar Gateway will provide around the clock relay whenever in view of the lander and rover, assumed to be ~85% of the time (Table 4.4-1). This may prove to be an unrealistic assumption, since the Gateway may need to independently communicate with several other spacecraft in different parts of the sky. The telecom rates described in this design may need to be increased to accommodate a smaller amount of time for communication. The Lunar Gateway operations infrastructure (Figure 4.4-1) will handle all space-ground communications, i.e., no budget for DSN or related ground stations was included.

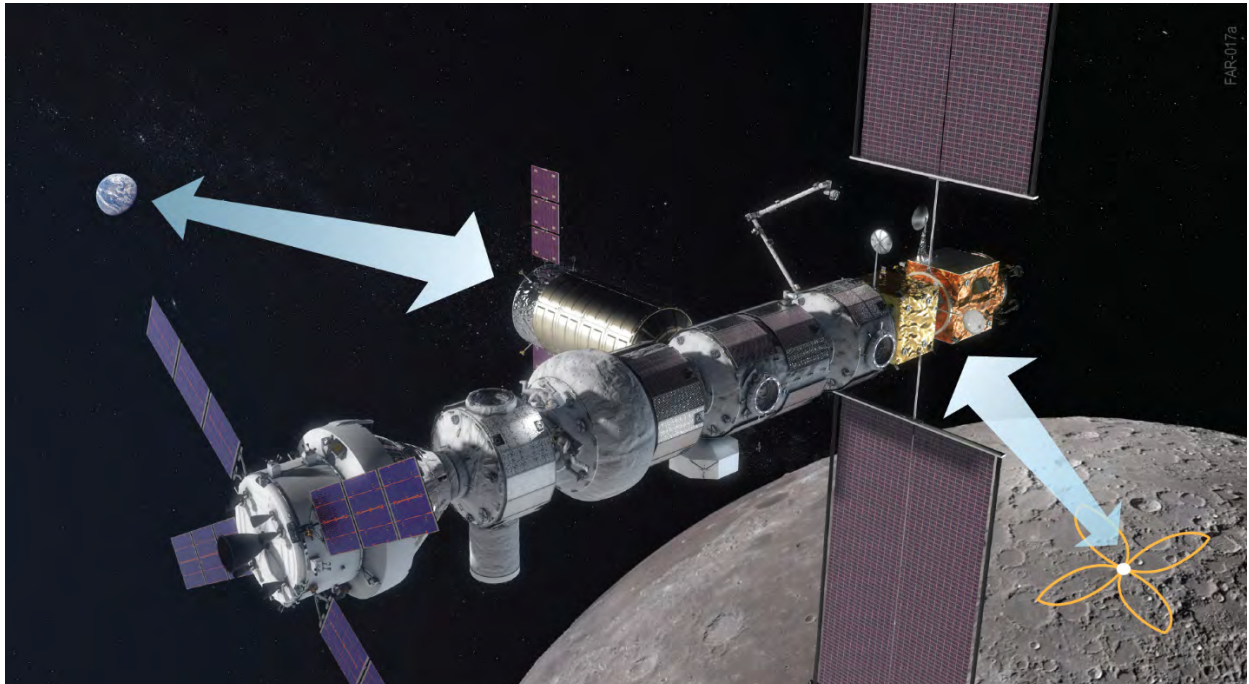

**Figure 4.4-1.** FARSIDE uses the Lunar Gateway, or similar Lunar asset, for communication with Earth.

### 4.4.1 *Mission Operation Phases*

The FARSIDE mission has four phases: Cruise, Initial Surface Ops, Spoke Deployment, and Science Operations, illustrated in Figure 4.4-2.

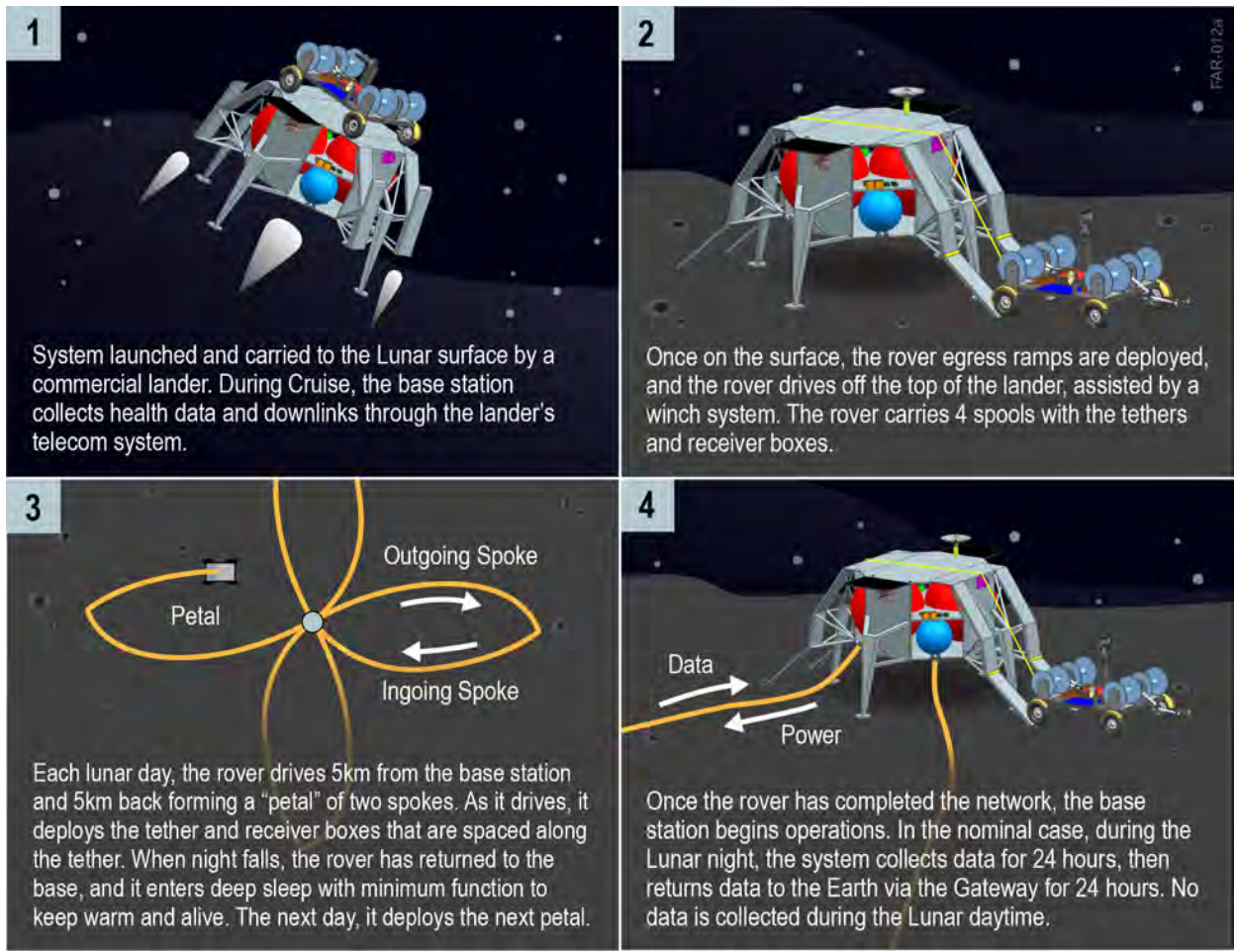

**1** System launched and carried to the Lunar surface by a commercial lander. During Cruise, the base station collects health data and downlinks through the lander's telecom system.

**2** Once on the surface, the rover egress ramps are deployed, and the rover drives off the top of the lander, assisted by a winch system. The rover carries 4 spools with the tethers and receiver boxes.

**3** Outgoing Spoke — Petal — Ingoing Spoke

Each lunar day, the rover drives 5km from the base station and 5km back forming a "petal" of two spokes. As it drives, it deploys the tether and receiver boxes that are spaced along the tether. When night falls, the rover has returned to the base, and it enters deep sleep with minimum function to keep warm and alive. The next day, it deploys the next petal.

**4** Data — Power

Once the rover has completed the network, the base station begins operations. In the nominal case, during the Lunar night, the system collects data for 24 hours, then returns data to the Earth via the Gateway for 24 hours. No data is collected during the Lunar daytime.

**Figure 4.4-2.** The FARSIDE mission has four phases: 1) Cruise; 2) Initial Surface operations, deploying the rover; 3) Spoke deployment—where the eight spokes are deployed as four petals; and 4) Science Operations.

## 4.5    Lander and Base Station

The 590 kg FARSIDE base station will use a commercial lander, assumed for the purposes of this study to be similar to the Blue Origin Blue Moon lander, which was scaled from photographs and renderings to determine the size. The electronics are fully dual string. The Base Station houses the FX correlator—each of the eight spokes feeds an F-engine board which receives the 32 optical signals (two polarizations per antenna node) and performs the frequency channelization in space qualified Field Programmable Gate Arrays (FPGAs). The output of those boards are corner-turned to X-engine boards that perform full cross-correlation of all antennas, also in FPGAs. The integrated output of the correlators is passed to the JPL-developed, radiation-tolerant, FPGA-based, Miniaturized Sphinx onboard computer and stored for relay to Earth. 192 GB of storage is provided, about three days at full rate. The Sphinx is also responsible for system management, engineering housekeeping telemetry, and control functions.

The rover rides on top of the lander, so that the entire assembly fits in the launch vehicle fairing. The rocket motor on the bottom of the lander prevents stowing the rover there, so a ramp system with a winch assist is used to get the rover from the cruise location to the lunar surface.

With no direct path to Earth, the Base Station communicates with the Lunar Gateway (or an alternative orbiting relay) using a conventional 10 Mbps RF link through a 90 cm gimbaled high gain antenna (HGA) during the 85% of the time that the Gateway is in view. Low gain antennas provide safe mode communications or when the HGA is unusable (i.e., if the position of the Gateway is not known). The baseline data rate for the array is ~6 Mbps, accommodated by the RF link. The telecom subsystem is essentially identical between the rover and the base station - both need to communicate to the gateway, and a relay would not work since the rover would travel farther away than the 1−2 km visible horizon.

### 4.5.1 Base Station Power

The baseline concept uses a pair of enhanced Multi Mission Radioisotope Thermoelectric Generator (eMMRTGs) with end of mission (EOM) power of 271W providing 30% contingency over normal requirements. Batteries provide additional reserve capacity to accommodate peak loads (311W for Science at night) during science data takes, and recharges during downlink times, when the antenna array and correlator are not powered on. The battery will discharge to approximately 60% during night-time science. The battery is 60 Ah, composed of two 30 Ah units, identical to the ones used on the rover. An active pumped fluid loop manages the waste heat.

The power system is conventional with the addition of a boost converter to take the Base Station bus voltage and boost it to the higher voltage (~300V) needed to feed the series strings of 16 antenna nodes. The power converter produces a constant current AC sine wave at approximately 20 mA. The frequency of the AC power supply is the subject of a future trade study with receiver node transformer size/mass and ripple requirements.

The power subsystem design and test campaign will consider EMI from the beginning of the project, since the switching rates and their harmonics likely fall within the sensitive bandwidth of the array. In keeping with standard JPL practice, all the converters will be synchronized to a common reference.

### 4.5.2 Base Station Thermal Design

Figure 4.5-1 shows the Base Station thermal subsystem. A zenith facing radiator rejects the excess heat from the eMMRTG, with variable conductance heat pipes used to adjust the amount of heat rejected to maintain acceptable temperatures between Lunar night and day. The avionics system has its own radiator with a pumped cooling system.

### 4.5.3 Base Station Electronics

The base station electronics (the design is shared with the rover) is based on the Sphinx boards developed at JPL [Imken, et al. 2017] for planetary missions which incorporate full redundancy into the system architecture. Custom boards will be designed to implement the FX correlator for processing the science data prior to sending it via the Gateway to Earth. Table 4.5-1 shows the Base Station requirements. The Base Station Electronics is approximately 23.6 kg mass and consumes ~68.4 Watts when performing science observing, but only ~10.4 watts when handling just communications and housekeeping.

**Table 4.5-1.** Base Station requirements.

| Data Volumes | Interfaces |
|---|---|
| Science data collection ~520 Gbits/Earth day (65 Gbytes) | Optical fiber interface for 128 receiver nodes |
| Storage capability ~195 GBytes (~3 Earth days of data) | Telecom (LVDS high rate data) |
| Base Station and Instrument health and status ~5 MBytes/day | Power/thermal |
| | Lander (during cruise) |

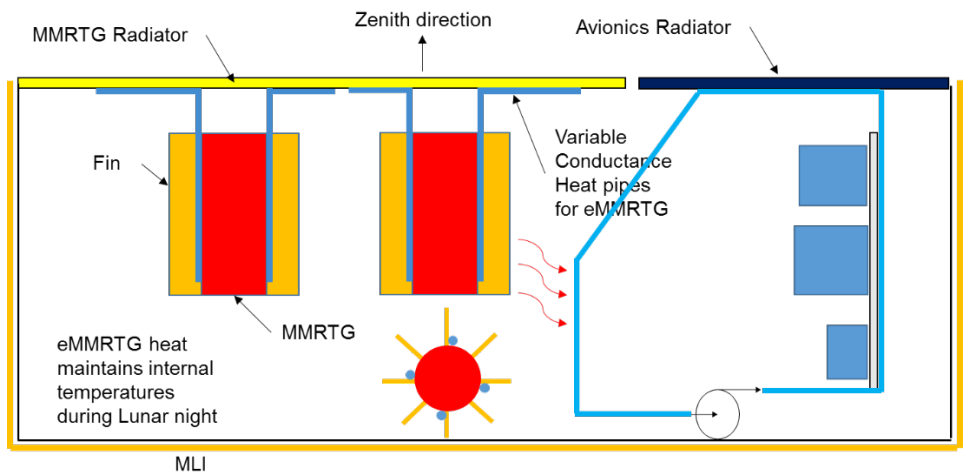

**Figure 4.5-1.** Base station thermal subsystem.

Figure 4.5-2 shows the Base Station electronics. Figure 4.5-3 shows the Sphinx Flight Computer and Sphinx radiation and memory performance.

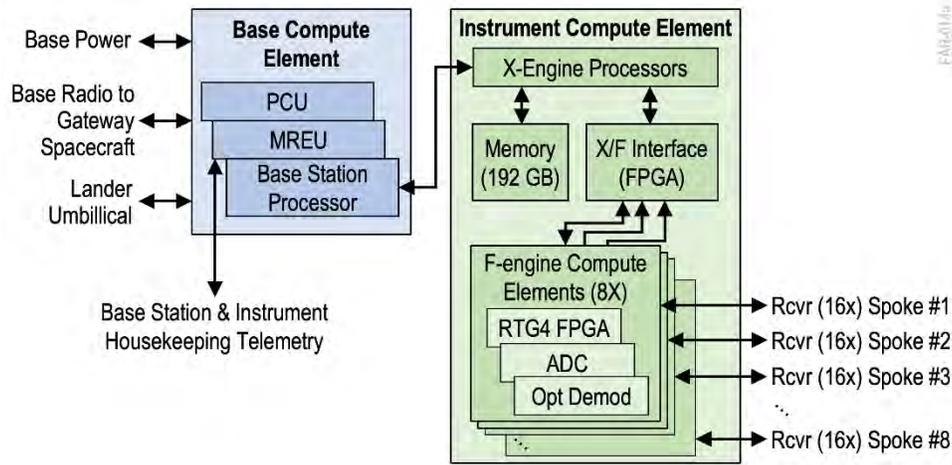

**Figure 4.5-2.** Base station electronics includes standard flight computer and specialized boards to perform radio astronomy calculations.

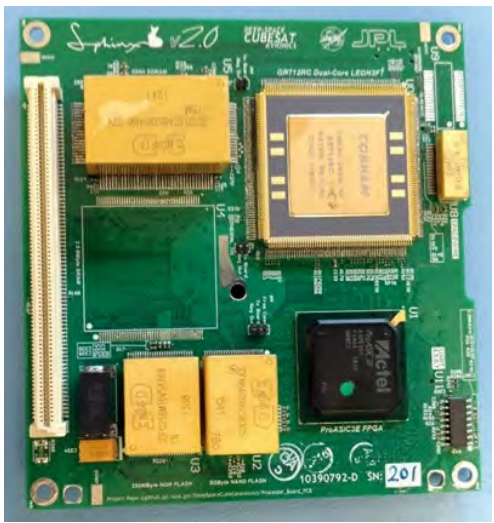

**Figure 4.5-3.** Sphinx Flight Computer and Sphinx radiation and memory.

**Radiation Performance:**
- TID: 30 Krads (Si)
  - GR712 Processor: 300 Krads (Si)
- LET:
  - 37 MeC/mg/cm2 for destructive SEs (SEL, SEB)
  - < 37 MeC/mg/cm2 for non-destructive SEs (SEU, SEFI, SET)

**Memory:**
- 97 KB on-chip memory (EDAC-protected)
- 256 MB SDRAM (EDAC-protected)
- 32 MB NOR Flash (EDAC-protected)
- 8 GB NAND Flash (EDAC-protected)

In addition to the digital signal processing, the base station electronics manage typical spacecraft communication, uplink and downlink, station housekeeping and fault management, during both cruise and after landing. The flight computer manages the MMRTG and fluid loop health and status as well as the instrument health and status.

The base station provides the digital signal processing for the 128 receivers on the 8 spokes, performing first frequency domain processing (the F of the FX correlator) and then performing cross correlation (the X). The frequency domain processing is done by F-engine Compute Elements (FCE) providing 16 dual polarization inputs with 80 MHz bandwidth. The FCE is implemented on 8 custom cards (one for each spoke) using rad hard FPGAs (Microsemi RTG4 or Xilinx Virtex 5) to process the data streaming in from Analog to Digital Converters (ADCs) that receive the 32 optical fiber signals. X-engine Compute Elements (XCE) perform the cross correlations across the entire 128 element array: F-engine buffers of ~8 MB each feed 8000 multiplier-accumulate. Bulk storage for processed science data is implemented with a SKR 192 Gbyte Flash system.

As with the power subsystem, radiated and conducted EMI is of some concern, which will be primarily addressed by packaging. Since the raw data from the receiver nodes is carried by optical fibers, there's no ingress pass there. Many EMI sources (CPU clock) that would be of concern in other systems will be less of a concern since they are at frequencies well above the FARSIDE instrument's maximum receive frequency.

### 4.5.4 Base station telecommunications

Both the base station and the rover share the same telecommunications subsystem design (Table 4.5-2) and general requirements:

- Support a two-way link with the Lunar Gateway through all landed mission phases.

- Support contacts with the gateway continuously.

  – (No communication from rover to Gateway is needed while rover is moving).

- Support a safe mode capability if HGA pointing is lost.

  We have assumed that the high rate links (10 Mbps return) will be at Ka-band and 16 kbps forward in Ku-band via a 0.9 meter antenna on a gimbal that can be pointed to the Gateway. Since there are dual redundant transmitter and receivers, no extra radio is needed for safe mode. All possible combinations of transmitters, receivers, and antennas are possible with the 2 RF transfer switches.

**Table 4.5-2.** Telecom parameters.

| Parameter | Uplink (forward) | Downlink (return) |
|---|---|---|
| Rate | 16 kbps | 10 Mbps |
| Safe Mode | 250 bps | 250 bps |
| Bit Error Rate | <1E-5 (uncoded) | |
| Frame Error Rate | | <1E-4 Rate ½ LDPC |

For the purposes of link budget calculations, we have assumed that the Gateway will have a 1.5 meter HGA for both Ka and Ku band, and that the maximum range to the Gateway is 75,000 km. It is likely that Gateway may have a larger receive (Ka-band) antenna, which would reduce the size requirement for both the base and rover HGA.

FARSIDE will use a version of the Electra Lite transponder which supports the Ku-band forward link and the Ka band return link. In keeping with the dual string, Class B architecture, there are two radios as shown in Figure 4.5-4. We have assumed that the radio will include a 5 W Ka-band Solid State Power Amplifier (SSPA - comparable to the Electra Lite including the UHF SSPA), although it may actually be a separate assembly for mechanical and thermal reasons, as in the Iris X-band radio.

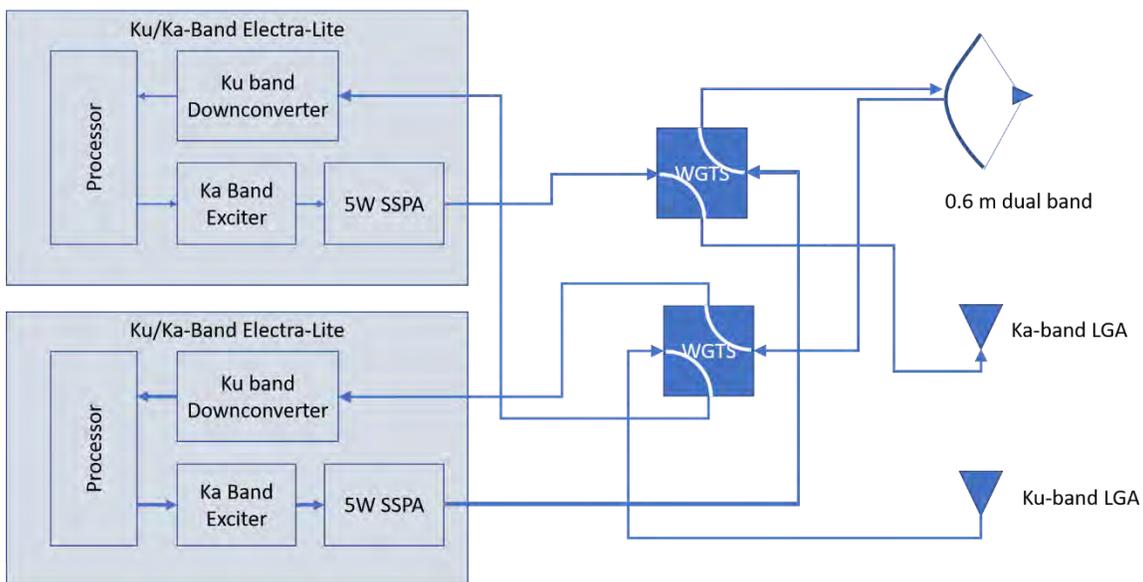

**Figure 4.5-4.** Rover and base telecom block diagram. HGA is 0.9 m for base station to support higher rates.

## 4.6    Rover Deployment

After landing, the rover would be deployed down a ramp to the Lunar surface, using a winch during the initial steep descent as shown in Figure 4.6-1. Ramps are provided on both sides for redundancy, in the event that the landing site makes it difficult to descend on one side or the other. Every 5 meters (about 3–4 minutes at the expected driving speed) a set of images would be collected and sent back to Earth, the teleoperators would confirm the next 5 meter driving increment, and the process repeats, roughly every seven minutes. An antenna node is deployed after every 3–4 hours of driving. Allowing one hour to deploy each antenna node, the trip out from the base to deploy 16 nodes would take 80 hours, and another 80 hours to return to base, deploying the second spoke. We have allowed one Lunar day (14 Earth days) to completely deploy one petal (two spokes) to allow for intervals when the Lunar Gateway is not in view, and for contingencies to route around obstacles. During the Lunar night, the rover goes into deep sleep mode. After 4 months, all petals would have been deployed.

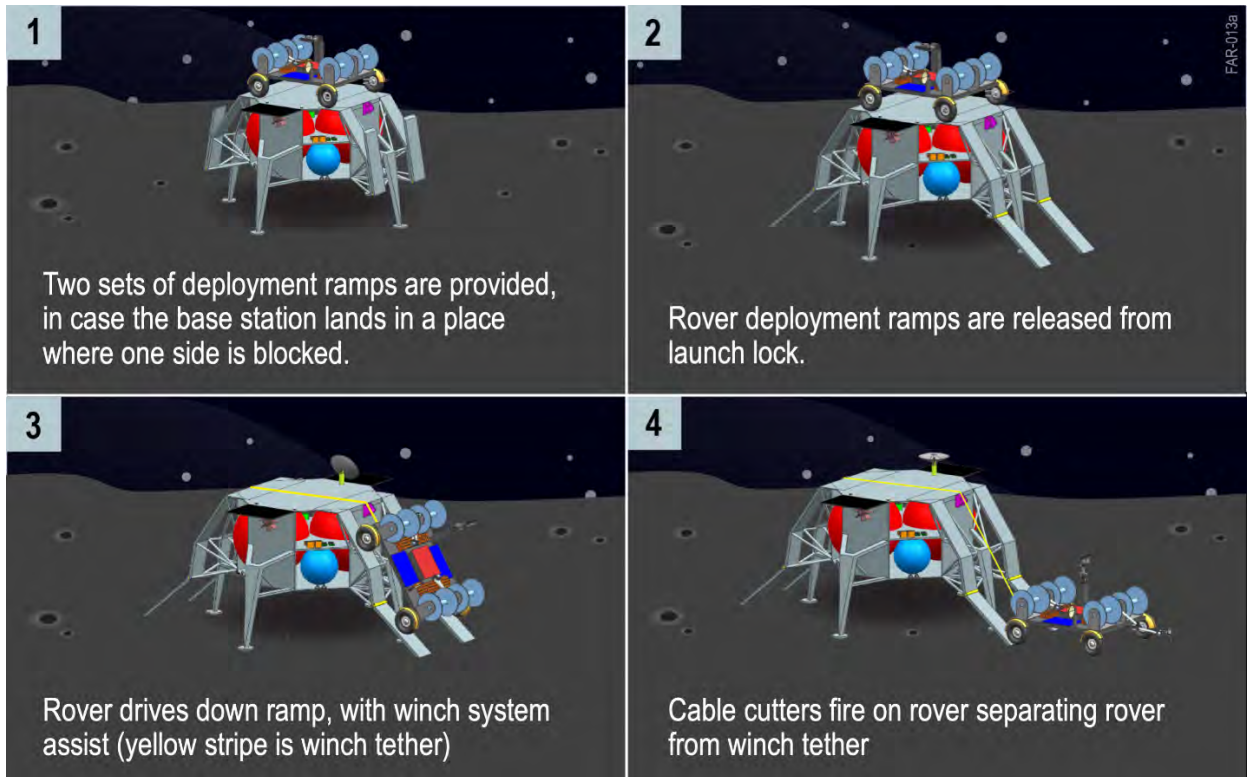

**Figure 4.6-1.** Conceptual sequence of the FARSIDE rover being deployed from the Base Station.

### 4.6.1    Rover Thermal Design

The 420 kg rover is based on the Apollo Lunar Roving Vehicle and includes 1.6 m$^2$ of solar array and a 30 Ah battery to stay alive during the Lunar night. Four RHUs are baselined to keep the avionics warm at night, and during the day, a 1.3 m$^2$ radiator would reject 180 Watts of heat during operations. Five thermal switches disconnect the radiator when temperatures drop after sunset. Figure 4.6-2 shows the Rover avionics module.

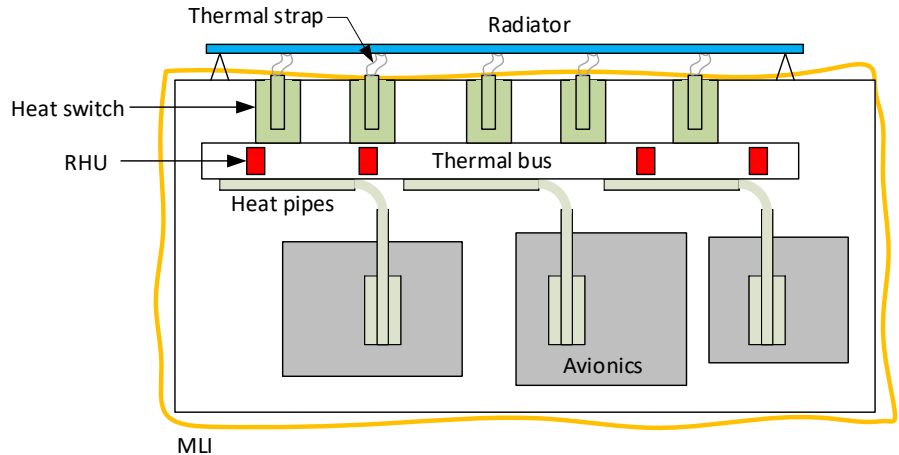

**Figure 4.6-2.** Rover avionics module includes RHUs to stay warm during the lunar night, and heat pipes to connect to radiator during day.

### 4.6.2    *Rover Power Subsystem*

At night, the rover shuts down, but when the sun hits the 1.3 m$^2$ (cell area) solar arrays on the rover, it autonomously powers on and turns on the radio in Rx mode, waiting for a command to power up the rest of the system. Figure 4.6-3 shows the required power and available solar power during a lunar day, deploying a single petal. The operations concept has the actual move and deploy starting two days after lunar sunrise. The 30 Amp Hr Lithium Ion battery provides power to keep the rover alive at night in a sleeping state, as well as operating the radio immediately after sunrise. The maximum depth of discharge in this scenario is 38%. A pair of the same battery system is also used on the Base Station, described later, so we only require one spare.

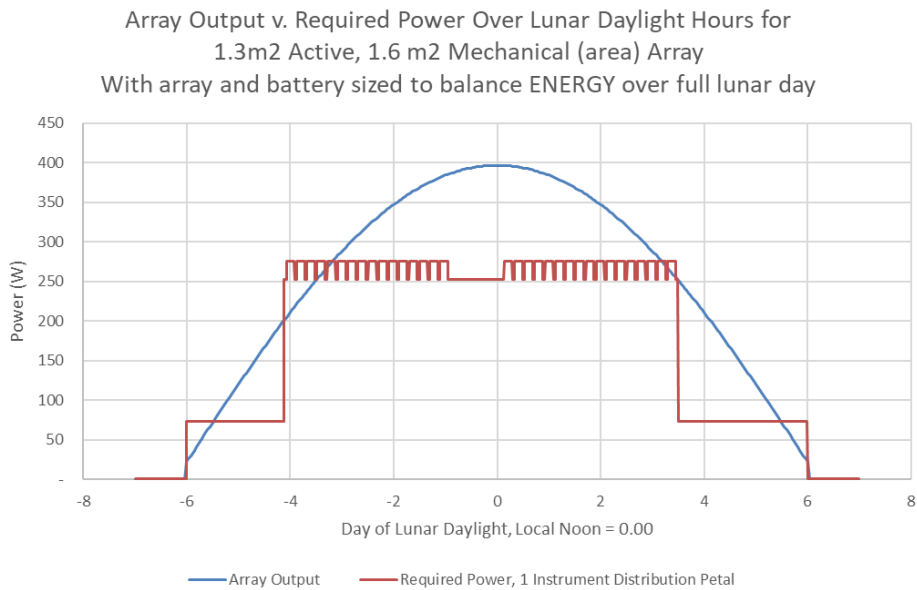

**Figure 4.6-3.** The required power and available solar power during a lunar day, deploying a single petal.

### 4.6.3    *Rover Electronics*

The rover carries the same flight computer and telecom package as the base station, as described in §4.5.3. The same JPL Sphinx flight computer (Figure 4.5-3) as the base station, connected to four hazard avoidance cameras and two navigation cameras on a mast which are used by human operators on Earth. Four motor controllers (derived from a current Europa Clipper design) in the motor control box are used for mobility and actuators. Figure 4.6-4 shows the Rover electronics. Since there is no intention to operate the array in science mode while the rover is turned on, the EMI characteristics of the rover are not a design driver.

### 4.6.4    Rover Telecommunications

The rover telecom electronics are identical to the base system, allowing sharing of spares during assembly, integration and test. The most important difference is that the rover has only a 5 Mbps maximum return data rate, because the HGA is 0.6 meters, 2/3 the size of the HGA on the base station (0.9 meters). This smaller antenna relaxes the pointing precision required and reduces the overall rover mass.

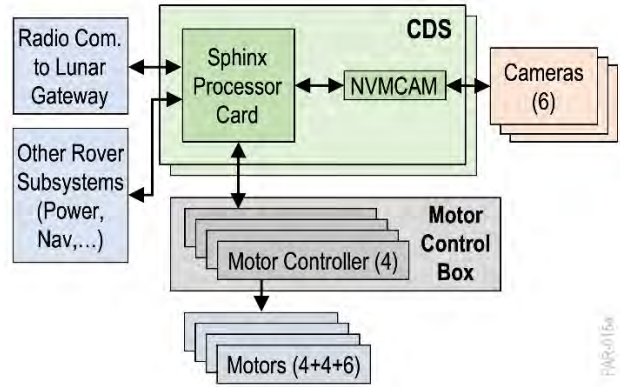

**Figure 4.6-4.** Rover electronics uses an identical flight computer as the base station, but includes motor controls and cameras.

The rover stops when communicating with the Gateway to reduce the maximum power load and to make it easier to control the pointing of the antenna. The operations concept is that the rover maintains position and attitude knowledge using odometry assisted by ground controller input. The rover stops, determines the look direction to the Gateway, points the HGA appropriately and transfers the data.

A low gain antenna provides a lower rate backup link if the rover cannot find the Gateway, or if position and attitude knowledge is lost.

### 4.7    Integration and Test

The FARSIDE probe base station, rover, and tethered receivers will be assembled at JPL and selected subcontractors. JPL environmental test lab and rover test yards are used for design, development and verification. Engineering model receivers and antennas will be tested in various outdoor facilities. Even though the soil properties are substantially different than Lunar regolith and the RF environment is much different than the relatively quiet Moon, the results from terrestrial tests can be modeled and verified.

The individual receivers will receive initial acceptance test during mass production manufacturing, and will be reverified after assembly into the tether, where final performance measurements are made using direct access connections to the antennas. The antenna element performance will not be verified on an individual basis, since they are electrically simple and identical.

A system testbed will be constructed at JPL and will evolve during the project development as designs and capabilities mature. The testbed will be used for initial runs of new procedures, fault diagnosis, and for trouble shooting in general. It will also be used for Operational Readiness Tests and for verification of proper handling of off-nominal conditions. The I&T element will be responsible for creating the testbed, assembling and maintaining it, certifying it where necessary. The system test bed also will provide sample inputs to the ground data system, using a simulator of the Gateway telecommunication link to be provided by others.

# 5 TECHNOLOGY MATURATION

The FARSIDE probe concept comes at a time of increasing interest in lunar exploration initiatives from NASA, international and commercial actors. This positions the mission to take advantage of key developments in lunar operations and infrastructure, such as the Lunar Gateway and commercial lunar payload services. With lunar initiatives constantly growing, FARSIDE comes at the opportune time with a unique architecture to achieve high-priority science objectives.

Technology maturation needs for FARSIDE can be broken into two bins: specific needs for the elements of the mission architecture (instrument receiver boxes, base station, and deployment rover) and programmatic needs for delivery to the lunar surface and operations. Current and future developments are expected to meet these needs in the mid-2020s, supporting a mission in the late 2020s or early 2030s.

## 5.1 Instrument

The FARSIDE receiver box instrument design uses mature technology with a simple implementation approach. An advanced reverse-operation thermal switch enables each node to survive the thermal environment of the lunar surface, fluctuating between extreme hot and extreme cold. The thermal switch passively decouples the radiator from the instrument box during the 14-day lunar night, allowing it to stay within allowable operating temperatures through the use of a single radioisotope heating unit (RHU). During the day, the radiator is recoupled, preventing the box from overheating. The thermal switch benefits from significant development at JPL to bring the technology to TRL 6.

Other elements of the receiver boxes require some degree of engineering development, such as the receiver antennas embedded within the science tether, but do not require significant technology development initiatives.

## 5.2 Base Station

The base station design uses high-heritage subsystem components whenever possible, with some new technologies which take advantage of current and continuing development initiatives by NASA. Because FARSIDE must survive and operate during the 14-day lunar night, a nuclear power source is needed. The design concept uses radioisotope thermal electric generators (RTGs), which have been employed in NASA science missions for decades. FARSIDE's power requirements demand a capability greater than the current generation of RTGs, the multi-mission radioisotope thermal electric generator (MMRTG), so the baseline design employs the enhanced version of this technology, known as the eMMRTG, which promises better efficiency and a higher power output. FARSIDE can also take advantage of NASA development on the Next-Generation RTG, which is envisioned to be a modular system providing a wide range of power output capabilities. Nuclear power sources are a critical asset to a wide range of NASA missions and the FARSIDE concept remains flexible to the specific implementation of a power source which NASA develops in the coming decade.

The large amount of science data that FARSIDE would collect requires a correlator architecture on the base station with significant computing capabilities. The baseline concept utilizes a custom F-engine compute element card on each of the eight instrument spokes. The compute elements take advantage of technology advances in radiation hardened FPGAs, such as the Microsemi RTG4 or Xilinx Virtex 5, both tested to over 100 krad total ionizing dose. The exact implementation of the custom correlator card is work that needs to be completed before the start of Phase A, requiring moderate technology and engineering development and testing.

## 5.3 Deployment Rover

The FARSIDE deployment rover builds on decades of NASA experience in operations on planetary surfaces. This includes lunar specific design inspiration from the Apollo Lunar Roving Vehicle, as well as remote operations experience from the multiple Mars rover missions that have been and continue to be flown since Mars Pathfinder/Sojourner in 1996. The design of the FARSIDE rover uses high heritage subsystem components, including guidance cameras and mechanical systems, with a power/thermal design that does not require a nuclear power source.

The unique complexities of the FARSIDE rover concept come from the operational requirements on the system. Specifically, the rover must drive over 40 km to deploy all of the receiver boxes, which would surpass any previous distance traveled for a planetary rover. The FARSIDE concept assumes teleoperation of the rover during the deployment phase by a human operations team on Earth. NASA has spent considerable effort to develop an understanding and technology for this type of operation for the Lunar Resource Prospector mission. FARSIDE can take advantage of this and other work by the NASA Ames Lunar Mission Planning and Operations Lab in lunar environment simulation and supervised autonomy to give confidence in the rover's ability to complete the deployment of the science instrument in the allocated time of four lunar days.

## 5.4    Commercial Lunar Lander

FARSIDE requires transportation to the lunar surface, assumed to be completed through the use of a commercial lunar lander. In the past decade, significant investments have been made by commercial companies to develop the capability to deliver payloads to the surface of the moon, with some companies now on the horizon of success. NASA shows strong support of these companies through the Commercial Lunar Payload Services Program (CLPS), which recently awarded the first contract to three companies for payload delivery with a launch target in 2021.

The current generation of CLPS landers have a modest payload capability on the order of 100 kg, with the goal of increasing the capability with the development of larger landers in subsequent years. The FARSIDE mass allocation of 1,750 kg fits well within the expected capabilities of future landers such as Blue Moon from Blue Origin. FARSIDE recognizes the government and commercial interest in lunar infrastructure that will bring about cost effective payload delivery to the surface of the moon in the next decade and takes advantage of that to enable this mission architecture.

## 5.5    Lunar Gateway

The far side of the moon does not have a line of sight to Earth, so a telecommunication relay capability is needed. The baseline FARSIDE concept assumes that this would be provided by the NASA Lunar Gateway, an orbiting station that is planned for launch in the early 2020's. The Gateway is expected to reside in a near-rectilinear halo orbit (NRHO), with view of the lunar far side during approximately 80% of its orbit. NASA has shown previous support for the use of the Gateway for science applications and infrastructure, such as telecommunications relay. The FARSIDE concept employs the Gateway for data return during the time that it is in view. Given NASA's current exploration goals with the Artemis program, the Gateway is a critical national asset that FARSIDE is able to take advantage of as an enabler of this architecture.

# 6    MANAGEMENT, HERITAGE, RISK, AND COST

FARSIDE is a collaboration between the University of Colorado Boulder (Principal Investigator Jack Burns), the California Institute of Technology (Co-PI Gregg Hallinan), Arizona State University (Co-Investigator Judd Bowman), NASA Goddard Space Flight Center (Co-I Robert MacDowall), the National Radio Astronomy Observatory (Co-I Richard Bradley), and the Jet Propulsion Laboratory. FARSIDE was initially developed, and funded, as part of the research program for the Network for Exploration and Space Science (NESS, https://www.colorado.edu/ness/), a NASA Solar System Exploration Research Virtual Institute (SSERVI) team. Paul Hertz selected FARSIDE for a Probe design study, recognizing the programmatic synergy between Decadal-level science and NASA's Artemis program of human/robotic exploration of the Moon. The FARSIDE team selected JPL as its partner for the design study and JPL's Team X as its concurrent design facility. The FARSIDE study was initiated in November 2018, well after the start of the competed Probe mission studies.

The FARSIDE Engineering Concept Definition Package was generated by Team X. The PI and Deputy PI participated in the Team X sessions, which consisted of an architecture study, an instrument study, and two mission studies for the lunar base station and deployment rover, resulting in a technically feasible point design that could be executed for a life cycle cost of ~ $850M, after allocating $150M for the launch vehicle. The science team was complemented by JPL team members from the SunRISE mission, recently selected by NASA for Extended Phase A study, because of the common frequency range and observational technique, and by Caltech personnel with expertise in astronomical signal processing.

## 6.1    Project Management Plan

FARSIDE will be implemented as a PI-led, science investigation. It is assumed that JPL will be the mission partner, responsible for Project Management, Project and Instrument Systems Engineering, Payload development and implementation, Flight System Management and Systems Engineering, Mission Operations, and Ground Data Systems. A JPL Project Scientist will support the PI, Deputy PI and science team, as is customary at JPL for a JPL-managed mission with a PI from the academic sector.

The high-level schedule is shown in Figure 6.1-1. A launch date of Jan 2028 is assumed, consistent with the assumptions in the Team X study. FARSIDE will be managed as a standard NASA flight project with typical mission phases, Phases A–D for development and Phases E–F for operations. The mission phase durations and concept of operations were developed as part of the study. The start of Phase A is assumed to occur in March 2022 to meet the proposed launch date. The first 7 months of Phase E are devoted to deployment, assembly, integration and testing, and calibration of the FARSIDE science instrument. The prime science operations phase is 60 months. Key review milestones are shown in the schedule: Preliminary Design Review (PDR), Critical Design Review (CDR) and System Integration Review (SIR).

FARSIDE depends on the implementation schedule for the Lunar Gateway and the evolution of Commercial Lunar Landers under NASA's Artemis program (https://www.nasa.gov/specials/artemis). The first phase of Artemis will culminate with humans landing on the moon in 2024. The second phase will establish a sustainable long-term presence by 2028. Blue Origins' Blue Moon lunar lander is designed to place scientific payloads, and eventually humans, on the surface of the moon, on the ambitious Artemis schedule (https://www.blueorigin.com/blue-moon, https://www.blueorigin.com/news/blue-origin-announces-national-team-for-nasas-human-landing-system-artemis). The first element of the Lunar Gateway, the Power & Propulsion Element (PPE), is already under development https://www.nasa.gov/feature/nasa-seeks-partnership-with-us-industry-to-develop-first-gateway-element, targeting a launch in 2022. These are also indicated on Figure 6.1-1. The early mission phases of FARSIDE are aligned with the Artemis program. System integration and test, which begins after the SIR, will start after the completion of the first phase of Artemis. A later FARSIDE start than that shown in the notional schedule only increases the likelihood that support services needed from the Lunar Gateway, possibly augmented with a network of relay orbiters to provide more complete lunar coverage, and a lunar lander such as Blue Moon that can be adapted as the foundation for the FARSIDE flight system, will be available. Technical and programmatic interfaces will be developed between the FARSIDE project, the Gateway, and the selected lunar lander subcontractor. NASA and JPL have ample experience implementing planetary missions on this scale with industry partners.

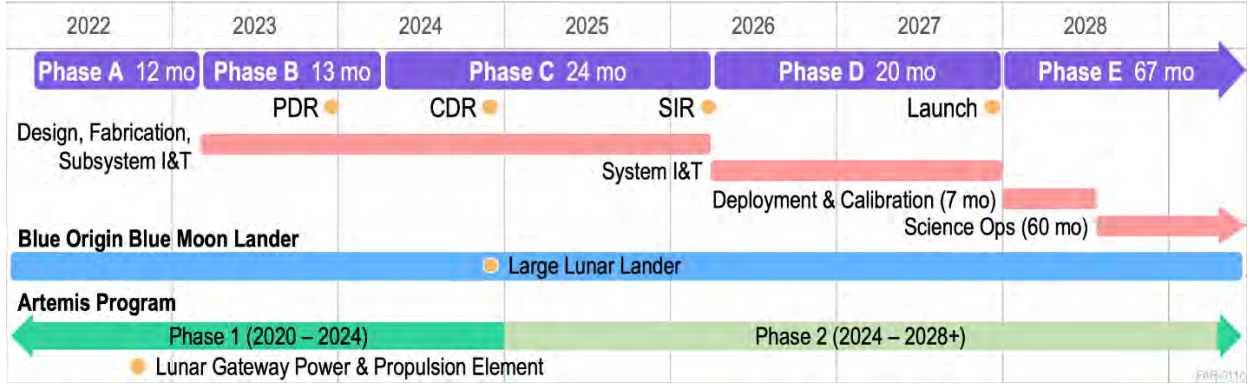

**Figure 6.1-1.** FARSIDE high-level schedule. The mission can be implemented in the 2025 – 2035 time frame and aligns with the development of the Lunar Gateway and lunar landers under the Artemis program.

## 6.2 Heritage and Risk

The FARSIDE science instrument, consisting of the receiver nodes and the F-X correlator in the base station, draws on significant heritage from low-radio frequency space missions, mature fiber-optic technology, and astronomical signal processing for Earth-based interferometer arrays. STACER antennas are flight-proven. Engineering development will be required to optimize the performance of the antenna elements in the science tether and the antenna-receiver signal coupling, and to balance performance over the frequencies required to meet science objectives. There is a lack of flight heritage for the instrument computing and interfaces, although JPL has ongoing strategic developments in radiation-hard, low size, weight and power single-board computing elements. Although the thermal design components have high Technology Readiness Levels (TRL), it may be challenging to obtain sufficient numbers of Radioisotope Heater Units (RHU) – the Department of Energy has not historically produced RHUs in this quantity.

The subsystems that comprise the base station and rover elements of the flight system all have significant heritage from successful JPL missions or NASA missions currently under development. The rover concept of operations is a risk area, because the concept assumes operator-in-the-loop, real-time driving operations with high interaction during the deployment phase. Additionally, the rover must travel ~ 50km on the lunar surface, perhaps surpassing the longest distance ever traveled by a solar-powered rover on a solar system body other than the Earth.

But the largest risks are programmatic – the dependence on NASA's commitment to the Artemis program and the technical evolution of the Lunar Gateway and the Blue Moon or comparable lunar lander.

## 6.3 Mission Cost

JPL's Integrated Design Facility (Team X) explored three architectures to initially assess technical feasibility and cost[1]. The $1B Probe-class mission nominal cost constraint was required for a successful candidate design. The selected architecture, consisting of 128 distributed receiver nodes, a central base station, and a single deployment rover, was first taken through an instrument study to develop the receiver box design and its key requirements and interfaces. This was followed by rover and base station studies, resulting in a technically feasible point design and the cost estimate summarized here. The Team X final study reports, which include full details for costing assumptions and basis of estimate, will be made available along with this report.

---

[1] The cost information contained in this document is of a budgetary and planning nature and is intended for informational purposes only. It does not constitute a commitment on the part of JPL and/or Caltech.

Table 6.3-1 shows the total mission cost, including Development (Phases A–D), Operations (Phases E-F), Launch Vehicle, and the cost of the eMMRTG and RHU elements. The basis of estimate for costing the eMMRTGs and RHUs is the guidance provided in the Discovery 2019 Announcement of Opportunity (AO). Reserves, 30% for Phases A–D and 15% for Phases E–F, meet JPL and NASA standards for competed missions during formulation. The cost is roughly consistent with the cost constraint within current uncertainties.

**Table 6.3-1.** Cost summary for FARSIDE.[1] Total mission cost by Phases A–D for development and Phases E–F for operations. For the eMMRTG and RHU costs, Team X adopted the cost guidance in the Discovery 2019 AO. No reserves were allocated for the launch vehicle or for the eMMRTG and RHUs.

| Cost Summary (FY2019$M) | | Team X Estimate | |
|---|---|---|---|
| | CBE | Reserve | Cost + Reserve |
| Total Cost | $1080M | 27% | $1330M |
| eMMRTG + RHU | $69M | 0% | $69M |
| Launch Vehicle | $150M | 0% | $150M |
| Development Cost (Phases A–D) | $800M | 30% | $1040M |
| Phase A | $8M | 30% | $10M |
| Phase B | $72M | 30% | $90M |
| Phase C/D | $720M | 30% | $940M |
| Operations Cost (Phases E–F) | $64M | 15% | $74M |

Costs were estimated using the standard NASA Work Breakdown Structure (WBS). Costs by WBS element for Development and Operations are shown in Table 6.3-2. Receiver node cost for the Theoretical First Unit (TFU) was derived from the NASA Instrument Cost Model (NICM), based on mass and power; the cost for additional receivers followed the Wright Learning Curve model [Wright 1936]. Both the rover and base station were costed as JPL in-house developments, for mission risk Class B with dual-string redundancy.

The deployment rover cost was derived using the Team X cost model for rovers, which draws heavily from JPL experience on the Mars Exploration Rover (MER) and Mars Science Laboratory (MSL) missions, including their actual costs, and from missions currently being designed at JPL, such as Mars2020, Europa Clipper, and Europa Lander. Cost drivers are the mobility system and deployment arm, the command & data handing system, rover software and a hazard avoidance capability consisting of a remote sensing mast and cameras.

The base station cost was derived using the Team X cost model for immobile, landed elements, based on preliminary design requirements. The key assumptions were that the descent vehicle would be the Blue Origins' Blue Moon commercial lunar lander, that lander sizing could be inferred from Blue Moon promotional material (https://www.blueorigin.com/blue-moon and https://www.blueorigin.com/news/blue-origin-announces-national-team-for-nasas-human-landing-system-artemis), and that all lander real estate could be used for configuring the integrated base station plus rover, with the exception of space used for lander propellant tanks. After rover deployment, the lander becomes the base station. Cost drivers are the mechanical structure, which includes the egress ramps, the winch and the rover deployment mechanism; the command & data handling system, which hosts all the digital signal processing to form the F-X correlator; base station software; and the power and telecommunication systems.

The other WBS elements were costed as percentages of these costs based on historical actual costs for NASA flight missions.

A number of cost risks were identified in the Team X studies. Little is known about the detailed design of the Blue Moon lander and its potential interfaces. Similarly, the capabilities of the Lunar Gateway are not well-known. A number of designs are new and may require additional development and testing beyond the scope of the Team X cost exercise for a pre-Phase A mission concept. Examples include the rover deployment from the lander, the science tether-to-base station interconnect, the concept of operations and timing for multiple (128) receiver box deployments, the base station computation elements and interfaces, the critical thermal design relying on RHUs to survive the lunar night, and a power system that can provide high voltage across the serial receivers. All require further study and are likely to claim resources from the 30% reserve allocated for Phases A–D.

**Table 6.3-2.** Pre-reserve cost breakdown by WBS, to WBS level 2. More detail, to lower WBS levels, is included in the Engineering Concept Definition Data Package, the Team X reports, in particular for the receiver boxes, the base station, and the rover.

| WBS Elements | Team X Estimate |
|---|---|
| Development Cost (Phases A–D) | $800.0M |
| 1.0 Project Management | $19.6M |
| 2.0 Project Systems Engineering | $21.6M |
| 3.0 Mission Assurance | $32.6M |
| 4.0 Science | $22.0M |
| 5.0 Payload System | $106.0M |
| 5.01 Payload Management | $1.2M |
| 5.02 Payload Engineering | $0.8M |
| Receiver Boxes (128 units) | $104.0M |
| 6.0 Flight System | $511.0M |
| 6.01 Flight System Management | $4.9M |
| 6.02 Flight System Systems Engineering | $35.3M |
| Base Station | $200.6M |
| Rover | $261.8M |
| 6.14 Spacecraft Testbeds | $8.6M |
| 7.0 Mission Operations Preparation | $28.2M |
| 9.0 Ground Data Systems | $26.9M |
| 10.0 ATLO | $31.1M |
| Operations Cost (Phases E–F) | $64.0M |
| 1.0 Project Management | $7.4M |
| 7.0 Mission Operations | $48.2M |
| 9.0 Ground Data Systems | $8.7M |